\documentclass[10pt]{article} 
\newtheorem{definition}{Definition}  
  
\newtheorem{proposition}{Proposition}  
  
\usepackage{vmargin}   
\setmarginsrb{3.2cm}{2.5cm}{3.2cm}{2.5cm}{1cm}{1cm}{1cm}{1.6cm}   
\usepackage{amsmath}   
\usepackage{amssymb}  
\usepackage[dvips]{graphicx}    
\usepackage{psfrag} 

\def\Clebphi#1#2#3#4#5#6{\left(\begin{array}{ll}     
\!#4 & \!#5 \\ \!\!\!#1 & \!\!\!#2 \end{array}\vline \begin{array}{ll} \!#3 \\ #6 \end{array}    
 \!\!\right)}  
\title{ {Cosmological Deformation of  Lorentzian Spin Foam Models }}  
 
\def\Clebpsi#1#2#3#4#5#6{\left(\!\begin{array}{ll}     
#6 \\ \!#3 \end{array}\vline \begin{array}{ll} #1 & \!\!\!#2 \\ #4 & \!\!\!#5 \end{array}\!\!    
 \right)}    

\def\sixj#1#2#3#4#5#6{\left\{\!\!\begin{array}{ll}     
#4 & \!\!\!#5 \\ #1 & \!\!\!#2 \end{array}\!\vline \begin{array}{ll} #3 \\ #6 \end{array}    
 \!\!\right\}}    

\def\Proof{\underline{\sf Proof}\;\;}
\def\onehalf{\frac{1}{2}}

\author{K. NOUI\thanks{e-mail:knoui@lpm.univ-montp2.fr}, Ph. ROCHE\thanks{e-mail:roche@lpm.univ-montp2.fr}{ }\thanks{Work supported by CNRS} \\  
 Laboratoire de physique math\'ematique et th\'eorique  \\  
 Universit\'e Montpellier 2, 34000 Montpellier, France.}  
\date{\today} 
 
\begin{document} 

\maketitle  

\begin{abstract} 
We study the quantum deformation of the Barrett-Crane Lorentzian spin foam model
which is conjectured to be the discretization of Lorentzian Plebanski model with positive cosmological constant and includes therefore as a particular sector quantum gravity in de-Sitter space.
This spin foam model is constructed using harmonic analysis on the quantum Lorentz group.
The evaluation of simple spin networks  are shown to be non commutative integrals over the quantum hyperboloid defined as  a pile of fuzzy spheres.
We show that the introduction of the cosmological constant, removes all the infrared divergences: for any fixed triangulation, the integration over the area variables is  finite for a large class of normalization of the  amplitude of  
the edges and of the faces.
 \end{abstract}

\section*{I. Introduction}
Based on a canonical quantization approach, the  loop quantum gravity program gives 
a  picture of the quantum geometry of space in terms of spin networks. 
Dynamic enters the theory only through  the Hamiltonian constraint which is still 
poorly understood.
 Spin foam models \cite{Bae} are  alternative way to give dynamic to spin networks and are candidates for the construction of a quantum theory of gravity in a covariant framework.
 Spin foam models are  attempts to evaluate  transition amplitude between spin network states. Therefore it extends  the structure of spin networks, which are one dimensional complexes,  to  two-complexes built from vertices, edges (colored by intertwining operators) and faces (colored by certain type of group representations).

 A spin foam model would ultimately describe the quantum geometry of space-time and any slice of it  would be  a spin network describing the space structure at a given time.

It is a major property of four dimensional pure  gravity  that it is a  constrained topological field theory: it can be expressed as a $BF$ theory with $SO(4)$ (resp. $SO(3,1)$) group for the Euclidean (resp. Lorentzian) signature, where the $B$ field is constrained to be ``simple'' and more particularly to be of the type $B=\star (e\wedge e).$

This property has given hope to construct spin foam models of quantum gravity by constraining spin foam models describing $BF$ theory.

Indeed Topological field theory of BF type can be exactly discretized with 
 spin foam models.
This is,  in three dimensions, the result of the work of  Ponzano-Regge \cite{PoRe} and in 4 dimensions the work of Ooguri \cite{Oo}. 

However without cosmological constant these models have  infrared divergences.
In the Euclidean signature case these previous spin foam models can be regularized by the addition of a cosmological term and one obtains, in three dimensions,  the Turaev-Viro model \cite{TV} and in four dimensions the Crane-Yetter model \cite{CY}. These spin foam models are built using   representation theory of certain quantum groups.

The Barrett-Crane model \cite{BC1} is a spin foam model which aim is to implement, in the Ooguri model, the constraint of simplicity on $B$.
This can be done very easily in the Euclidean case by restricting representations on the faces to be ``simple'' and by coloring the edges with a very peculiar intertwiner, known in the litterature as the Barrett-Crane intertwiner.
Soon after this proposal these authors gave a similar proposal \cite{BC2} in the Lorentzian case.

These models, which gave an impetus for a large amount of work on spin foam models of constrained $BF$ theory, have several drawbacks:

1) The weight of the vertices  is well defined (a 10j symbol) but the weights on the edges and faces  are not fixed and are still a matter of debate \cite{BCHT}. This issue is very important for proving the finiteness of the amplitude after integrating over the area variables for a fixed 2-complex
 \cite{Pe,CPR}.

2) A meaning to the  sum over 2-complexes  has to  be done in order to  compute an 
amplitude, and up to now, there is no result
 on this important issue in the 4 dimensional case. 
Note however the recent work of \cite{FL} in the case of three dimensional Riemmannian  quantum gravity.

3) The simplicity constraint does not imply that the $B$ field is in the gravitational sector, and the relative weight of the gravitational sector compared to the other seems to be small \cite{Re1}.

Our work aims at defining and studying the Barrett-Crane model in the case of the quantum Lorentz group $U_q(so(3,1))$ for $q$ real. This is possibly interesting 
for different reasons.

The first one is that one would like to obtain spin foam models for describing quantum gravity with a positive cosmological constant. 
This is important for the study of   quantum gravity in de Sitter space \cite{Sm}. 

The second one is that the use of   quantum groups enhance convergence, particularly in the infrared. 
In \cite{BCHT} three spin foam models are analyzed in the Riemannian case. They differ only by the choice of the weight on the edges and on the faces.
We will show that for any $2$- complex dual to a triangulation of the manifold, the integration over the coloring (i.e area variables ) are finite for the quantum deformations of these three models.

The third one is that, if the sum over 2-complexes can be defined, it is only in the presence of a cosmological constant.

Our work is divided in four sections.
In section II we recall elementary facts on spin foam models.
In section III we give a construction of the quantum hyperboloid and of the quantum Lorentzian Barrett-Crane intertwiner.
Section IV is devoted to the study of quantum simple Lorentzian spin networks.
Section V is an analysis of the cosmological deformation of the Lorentzian 
Barrett-Crane model. 
We will prove the infrared finiteness property in this section.

\section*{II. Spin foam models, quantum gravity and cosmological constant}

A large class of Topological Quantum Field Theories  can be exactly discretized by spin foam models. For a review on this subject, see \cite{Bae,FK1}. This class of theories  includes $BF$ theory in any dimension and $BF$ theory with cosmological constant in three and four dimensions. The Barrett-Crane model is a spin foam model which is advocated to be a discretization of the quantization of Plebanski theory  in four dimensions. The Euclidean version has been defined in \cite{BC1} and the Lorentzian one was studied in \cite{BC2}. In this last paper, the authors have also mentioned the idea to extend their model to the quantum deformation of the Lorentz group as a possible regularization.

 After an overview of spin foam models (Lorentzian and Euclidean) in 3 and 4 dimensions, we recall the links between quantum spin foam models and $BF$ theory with cosmological constant. We then explore the relations between quantum deformation of Lorentzian Barrett-Crane model and quantum gravity in the presence of a cosmological constant.

In the sequel we will consider a Riemannian (resp. Lorentzian) manifold. 
We will denote $\sigma=\pm 1$ the sign of the metric.  We will denote $\eta=diag(\sigma,1,...,1)$ and $SO(\eta)$ the isometry group of $\eta.$ 

The  Einstein-Hilbert action on a $D$ dimensional oriented  Riemannian (resp.Lorentzian) manifold with cosmological constant $\Lambda$ is  defined by
\begin{eqnarray} \label{actionofgravity}
S_{EH}[g]= \frac{1}{G} \int_{M} d^D x \;  \; ({ R}[g]-2 \Lambda)
 \sqrt{\vert  g\vert}.
\end{eqnarray}
 Let us introduce an orthonormal triad $e_I=e_I^{\mu} \partial_{\mu}$ and a $so(\eta )$ connection $\omega_{\mu}{}^I_J$. The metric can be expressed in term of the cotriad $e^I$ as $g_{\mu \nu}= e^I_{\mu} e^J_{\nu} \eta_{IJ}$ and the action (\ref{actionofgravity}) is rewritten as the Palatini action
\begin{eqnarray}
S_P[e,\omega] = \frac{1}{G} \int_{ M} d^D x \; \vert e \vert \; (R_{\mu \nu}^{IJ}[\omega]\;  e_I^{\mu} \; e_J^{\nu}-2\Lambda)
\end{eqnarray}
where $e= det(e^I_\mu)$. 
This action can be conveniently written in term of forms, as:

\begin{equation}
S_P[e,\omega]=\frac{1}{G} \int_{ M} 
 (R^{IJ}[\omega]\wedge {\star}( e_I\wedge  e_J)-2\Lambda {\star}
1)
\end{equation}
where the $\star$ is the Hodge operator.
 From the orthonormality of $e_I$, we have:
 \begin{equation}
\star (e^{I_1}\wedge\cdots\wedge  e^{I_r})=\frac{s(e)}{(D-r)!}\epsilon^{I_1\cdots I_r}_{I_{r+1}\cdots I_{D}}e^{I_{r+1}}\wedge\cdots\wedge  e^{I_D}
\end{equation}
where $s(e)$ is the sign of $e.$

The Palatini action can be recast, in any dimension as:
\begin{equation}
S_P[e,\omega]=\frac{1}{(D-2)!G} \int_{ M} 
s(e) \epsilon_{I_1\cdots I_{D}}(R^{I_1I_2}\wedge
e^{I_3}\wedge\cdots\wedge e^{I_D} -
\frac{2\Lambda}{D(D-1)} e^{I_1}\wedge\cdots\wedge e^{I_D}).
\end{equation}

	\subsection*{II.1. Spin foam models in 3 dimensions}

The very first spin foam model was introduced by Ponzano and Regge \cite{PoRe} as an attempt to define a discrete version of three dimensional Riemannian quantum gravity. The modern understanding of this construction follows the next  lines of arguments. 

The action of $BF$ theory in $3$ dimension for the $SO(\eta)$ group is  given by:
\begin{equation}
S_{BF}=\frac{1}{G} \int_{ M} 
\epsilon_{IJK} R^{IJ}\wedge
e^{K},
\end{equation}
as a result in the case where $s(e)=1$ this is exactly the action of Palatini.

The partition function of $BF$ theory is 
\begin{eqnarray} 
{\cal Z}_{BF} = \int {\cal D}e {\cal D} \omega \exp \left( \frac{i}{\hbar} S_{BF}[e,\omega] \right).
\end{eqnarray}

The action, being linear in  the  $e^I$, we can integrate over the $e^I$ and we get:
\begin{eqnarray} 
{\cal Z}_{BF} = \int {\cal D} \omega
\prod_{x\in M}\delta(R^{IJ}[\omega](x)).
\end{eqnarray}
This partition function is therefore the ``volume'' of the space of flat connections on $M.$ It can be exactly computed by the following discretization in the Riemannian case.
 We choose any triangulation ${\cal T}$ of the manifold: to each edge $\epsilon$ we associate an irreducible representation $j_\epsilon$ of $SU(2)$. The Ponzano-Regge partition function is then given by  
\begin{eqnarray} \label{PRsum}
{\cal Z}_{PR} ({\cal T}) = \sum_{\{j_\epsilon\}} \prod_\epsilon dim(j_\epsilon) \prod_t W_t\;,
\end{eqnarray}
where the sum runs over all irreducible representations of $SU(2)$. We have assigned the weight $dim(j_\epsilon) = 2j_\epsilon+1$ to each edge $\epsilon$ and the weight $W_t= \exp(i\pi\sum_{\epsilon\in t}j_{\epsilon})(6j)_t$ to each tetrahedron $t$ colored with the  six representations labelling the edges of $t$. Note that the $6j$ we are using in this expression is invariant under the group of symmetry of the tetrahedron.
The length of the edge $\epsilon$ in natural units is $l(\epsilon)=l_P dim(j_\epsilon)$, where $l_P = G \hbar$ is the Planck length.
A simple explanation of this assertion comes from the classical limit of Ponzano-Regge sum: we let $l_P$ tends to zero while letting $l_P dim(j_\epsilon)$ fixed to a classical length $l(\epsilon)$.
Thus the weight of a tetrahedron $t$ behaves as \cite{PoRe}:
\begin{eqnarray} \label{PRasymptotic}
(6j)_t \sim  \frac{1}{ \sqrt{12 \pi  V[t] }}\cos(\frac{S_R[t]}{\hbar}+\frac{\pi}{4})
\end{eqnarray}
with $V[t]$ the volume of the tetrahedron and the Regge action \cite{PoRe}, given by 
\begin{eqnarray}
S_R[t]=\frac{1}{G}\sum_{\epsilon\in t}\theta_\epsilon \; l(\epsilon) \; ,
\end{eqnarray}
is a discretization , for $\Lambda=0$,  of (\ref{actionofgravity}).
The partition function  (\ref{PRsum}) can also be expressed as a sum over spin foams lying in the dual 2-skeleton of the given triangulation: in this way, the Ponzano-Regge model appears to be  a spin foam model. Note that the sum can be shown to be triangulation independent (because of the Biedenharn-Elliot relation)  and thus it defines a topological  invariant of the manifold $M$.

\medskip

 The Ponzano-Regge sum, as well as the partition function of $BF$ theory,  differs from the partition function of $3$ dimensional gravity.  Indeed the weight of a  configuration of $e$ with $\text{det}\; e>0$  differs  from a weight of $e$ with $\text{det}\; e<0$ in $BF$ theory whereas it is the same in the action of gravity. This problem is too much understated in the literature but is well 
addressed in \cite{FK2}. 
The comparison between the two theories is still not very well  understood.

The Ponzano-Regge sum suffers from an infrared divergence, i.e when the variables $j_\epsilon$ become large. Several possible regularization have been introduced,
 among them the Turaev-Viro one is the most natural and effective. It consists  
 in replacing the classical Lie algebra  $su(2)$ by the corresponding quantum group $U_q(su(2))$ where $q$ is the   root of unity $q=\exp(i\frac{2\pi}{k+2})$. The edges are now labelled by irreducible representations of $U_q(su(2))$ and the sum becomes:
\begin{eqnarray}\label{sumTV}
{\cal Z}_{TV} ({\cal T}) = K^{\vert V\vert} \sum_{\{j_\epsilon\}} \prod_\epsilon dim_q(j_\epsilon) \prod_t W_t^q\;,
\end{eqnarray}
where  $dim_q(j_\epsilon)$ is the quantum dimension of the representation $j_\epsilon$, and $W_t^q=\exp(i\pi\sum_{\epsilon\in t})(6j_q)_t$ , where $(6j_q)_t$ is the quantum $6j$ symbol associated to the tetrahedron $t$, $
\vert V\vert $ is the number of vertices of ${\cal T}$ and 
$K=-\frac{(q^{1/2}-q^{-1/2})^2}{2(k+2)}>0$.

Note that  the number of irreducible representations of  $U_q(su(2))$ with non vanishing q-dimension is finite when $q$ is a root of unity, therefore 
 the sum (\ref{sumTV}) is a finite sum.

The Turaev-Viro model is an exact discretization of $BF$ quantum theory with a cosmological term, whose action is:
\begin{eqnarray}
S_{BF,\Lambda}[e,\omega]= \frac{1}{G} \int_{ M}  \epsilon_{IJK} \; (R^{IJ}[\omega]-\frac{\Lambda}{3} e^I\wedge e^J) \ \; \wedge \; e^K.
\end{eqnarray}
When $e>0$, this action is the same as the Euclidean Einstein-Hilbert action with   cosmological constant $\Lambda>0,$
and the  level $k$ is related to the cosmological constant by $k= \pi\frac{l_c}{l_P} $ where $ l_c= \frac{1}{\sqrt{\Lambda}}
$ is the cosmological length.

Note that the partition function ${\cal Z}_{TV}$ is unchanged when one turns  $q$ into  $q^{-1}$ because of the invariance of the quantum dimension and the quantum $6j$-symbol in this transformation. This property corresponds in the $BF$ theory to the invariance  of the partition function ${\cal Z}_{BF,\Lambda}$ when one tuns  $e_I$ into  $-e_I$ and conjugate $i$ into  $-i$. 

There are, up to our knowledge, different ways to understand that the Turaev-Viro sum is related to $BF$ theory with cosmological term and that quantum groups necessarily  appear in the quantization process.

The first one, well exposed for example in \cite{FK2} uses the equivalence, due to Witten \cite{Wi1},  between 
$BF$ theory with positive cosmological term and Chern-Simons theory with group
$Spin(4)=SU(2)\times SU(2).$ 
In order to show that the Turaev-Viro sum is the partition function of $BF$ theory with positive cosmological term one needs an additional result,  shown by mathematicians, which is that the 
Turaev-Viro sum of a manifold $M$ is the modulus square of Reshetikhin-Turaev invariant of $M$ for the group $SU(2).$ The equality between Reshetikhin-Turaev invariant of $M$ and the partition function of Chern-Simons of M implies therefore the equality between the Turaev-Viro sum and the partition function of BF with cosmological constant as well as the relation between $q$ and $\frac{l_c}{l_P}.$

An alternative  way consists in studying the asymptotic  behavior of the quantum $6j$-symbol in the regime where
\begin{eqnarray} \label{3dasymptotic regime}
l_P \ll l_\epsilon \ll l_c\;\;.
\end{eqnarray}
This regime  corresponds to the classical limit with a finite cosmological constant: physically, the length are bounded by $l_c$ because of the presence of a cosmological horizon. The asymptotic value of the quantum $(6j_q)_t$ has been well understood in \cite{MT} where, after reintroducing the natural units, the authors obtain:
\begin{eqnarray} \label{TVasymptotic}
(6j_q)_t \sim \frac{1}{ \sqrt{12 \pi  V[t] }} \cos(\frac{S_{R,\Lambda}[t]}{\hbar}+\frac{\pi}{4}),
\end{eqnarray}
where $V[t]$ is the volume of the tetrahedron and the Regge action in presence of a cosmological constant is
\begin{eqnarray}
S_{R,\Lambda}[t]=\frac{1}{G}\left(\sum_{\epsilon\in t}\theta_\epsilon \; l(\epsilon) \; - \frac{1}{l_c^2} V[t]\right)\;\;.
\end{eqnarray} 

These two previous arguments are rather indirect. A very clear understanding of the appearance of quantum groups in the quantization of $BF$ theory with cosmological term is obtained using  the combinatorial quantization of Chern-Simons theory as first understood in \cite{FR}.

\medskip

There are recent attempts to formally define 3-dimensional Lorentzian spin foam models, with possible applications to 2+1 quantum gravity.

An analogue of the Ponzano-Regge sum can be straightforwardly defined.
One begins with the partition function of $BF$ theory with $SO(2,1)$ group and applies the  same method of integrating the $e$ field and  replacing the delta function   by its decomposition into characters of unitary representations of $SO(2,1)$. One is therefore led to define a Ponzano-Regge model where the representations coloring the edges belong to the principal series and to the discrete series. The weight of the tetrahedron is given as the $6j$ of these representations \cite{Fr,Da}.
This subject is still in its infancy, and important properties such as the
study of particular regularization of this $SO(2,1)$ Ponzano-Regge sum and the  proof of the asymptotic behaviour of the $6j$ symbols are still unknown. 

It is expected that quantization of  $so(2,1)$ $BF$ action with positive cosmological term is  described by a Turaev-Viro model associated to  $U_q(su(1,1))$ for q real with  $q=\exp(-l_P/l_c).$
Due to the equality between the $BF$ action with positive cosmological term and the Chern-Simons action associated to  the Lie algebra $so(3,1)$, there should exist  a relation between the would be Turaev-Viro sum with $U_q(su(1,1))$ and the would be  Reshetikhin-Turaev invariants for the quantum algebra  $U_q(so(3,1)).$ To be defined, these invariants, as well as the expectation value of observables,  need regularization and up to now the litterature is empty on this subject. Note however our work \cite{BNR}, where
 we quantize Chern-Simons theory with $SO(3,1)$ group using the program of combinatorial quantization of Hamiltonian Chern-Simons theory.

	\subsection*{II.2. Spin foam models in 4 dimensions}

		\subsubsection*{II.2.1 BF theory and Crane-Yetter invariants}

The action of $BF$ theory in $4$ dimensions in the case of a simple compact Lie group $G$ is 
\begin{eqnarray} \label{BFaction}
S_{BF}[B,A] = \int_M Tr\left( B \wedge F(A)\right)
\end{eqnarray}
where the field $B=B_{\mu \nu}
dx^{\mu} \wedge dx^{\nu}$ is a $\mathfrak{g}$-valued two-form, $F$ is the curvature associated to the $\mathfrak{g}$-connection $A$ and $Tr$ denotes the Killing form on the Lie algebra. 
\medskip

\begin{figure} 
\psfrag{a1}{$j_1$}
\psfrag{a2}{$j_2$}
\psfrag{a3}{$j_3$}
\psfrag{a4}{$j_4$}
\psfrag{a5}{$j_5$}
\psfrag{a6}{$j_6$}
\psfrag{a7}{$j_7$}
\psfrag{a8}{$j_8$}
\psfrag{a9}{$j_9$}
\psfrag{a10}{$j_{10}$}
\psfrag{b1}{$i_1$}
\psfrag{b2}{$i_2$}
\psfrag{b3}{$i_3$}
\psfrag{b4}{$i_4$}
\psfrag{b5}{$i_5$}
\centering 
\includegraphics[scale=0.6]{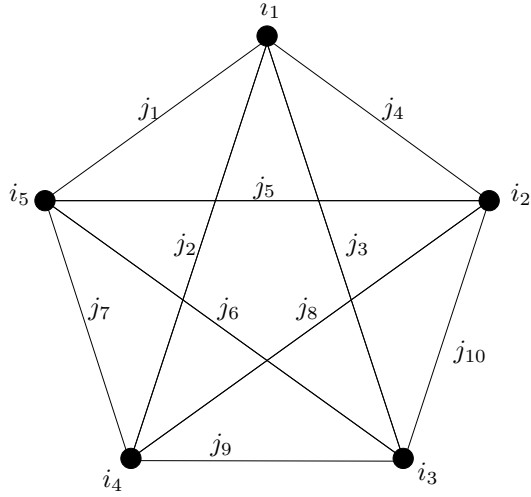} 
\caption{Diagramatic representation of the $15j$-symbol.} 
\label{15jsymbol1} 
\end{figure}

This defines a topological theory which has been first studied in 
\cite{Ho}. In the case of $G=SU(2)$, which is the case that we consider in this section,
Ooguri \cite{Oo} was the first to provide an exact  discretization of $BF$ theory in 4 dimensions through the construction of a spin foam model. Let $M$ be a four dimensional oriented compact smooth manifold. Let ${\cal T}$ be any triangulation of this  manifold and $\Delta_n$ the set of $n-$ simplices of
 ${\cal T}$. Let ${\cal J}$ be the set of  irreducible representations of $G$. If $j\in  {\cal J}$, we denote $V_j$ the vector space on which the representation $j$ acts. To each face $f$ of ${\cal T}$  we associate an element $j_f$ of ${\cal J}.$ To each 
tetrahedron $t$ we define $I_t$, to be the space of $su(2)-$intertwiners $\phi:\bigotimes_{f\in t}V_{j_f}\rightarrow {\mathbb C}.$ $I_t$ is endowed with the Hermitian form $<\phi,\psi>=tr(\phi\psi^{\dagger}).$
In the $SU(2)$ case an orthonormal basis of $I_t$ is labelled by elements of ${\cal J}$ as well. To each tetrahedron we associate an element $i_t$ of this orthonormal basis or equivalently  an element $j_t \in {\cal J}.$  The Ooguri partition function is defined as:
\begin{eqnarray} \label{Osum}
{\cal Z}_O \left({\cal T} \right) = \sum_{\{j_f\}}\sum_{\{j_t\}}
 \prod_{f\in \Delta_2} dim(j_f)
 \prod_{t\in \Delta_3} dim(j_t)^{-1} 
\prod_{s\in \Delta_4} (15j)_s\;\;,
\end{eqnarray}
where the so-called $15j$-symbol $(15j)_s$ associated to each 4-simplex $s$ of the triangulation is constructed from the 10 representations labelling its 10 faces and the 5 representations labelling its 5  tetrahedra (figure \ref{15jsymbol1}). 

For the same reason as for the Ponzano-Regge model, the sum (\ref{Osum}) is in general  divergent.
The Crane-Yetter model \cite{CY,CKY1} is a regularization of (\ref{Osum}). It is the 4-dimensional analogue of the regularization of Turaev-Viro of the Ponzano-Regge sum and it is an exact discretization of $BF$ theory with cosmological term
whose action is:
\begin{equation}
S_{BF,\Lambda}= 
\int_M Tr\left( B \wedge F(A)\right)+\Lambda
\int_M Tr\left( B \wedge B\right).
\end{equation}

The regularization consists  in replacing the classical Lie group by the quantum group $U_q(su(2))$ where $q$ is taken to be a root of unity. However, because it is impossible to have a projection of the 4-simplex on a plane without having at least two edges which intersect, it is important to be careful on the definition of the $15j$ symbol and to correctly take account of the crossing.. This  issue does not appear in the Turaev-Viro construction.
The Crane-Yetter construction  was generalized to the framework of
 2-spherical categories in \cite{Ma}. We will refer to this work and use its notations  because the construction is cristal-clear there.

 We take the quantum algebra $U_q(su(2))$ for $q$ root of unit.
We pick any   total order on the set of  vertices of the triangulation ${\cal T}$. This order induces, by lexicography, a total  order on the set of faces, on the set of  tetrahedra and on the set of 4-simplices.
To each face $(ijk)$ we associate a representation of $U_q(su(2))$ and we denote by $V(ijk)$
its associated vector space. Let $s$ be a $4$-simplex of ${\cal T}$, $s=(ijklm),$ 
\begin{eqnarray*}\label{boundary4simplex}
\partial s= (jklm)-(iklm)+(ijlm)-(ijkm)+(ijkl)\;\;.
\end{eqnarray*}    
To each tetrahedron, appearing with positive sign in $\partial s$, for example $(ijkl)$,  we associate the space $H^+(ijkl)$ of  intertwiners $\alpha_{ijkl}^+$
\begin{eqnarray} \label{+intertwinerijkl}
\alpha_{ijkl}^+ \; : \; V(ijl) \otimes V(jkl) \longrightarrow V(ikl) \otimes V(ijk).
\end{eqnarray}
To each tetrahedron appearing with negative sign in $\partial s$, for example $-(iklm)$, we associate the space  $H^-(iklm)$ of intertwiners $\alpha_{iklm}^-$
\begin{eqnarray} \label{-intertwinerijkl}
\alpha_{iklm}^- \; : \; V(ilm) \otimes V(ikl)
 \longrightarrow V(ikm) \otimes V(klm).
\end{eqnarray}

\begin{figure} 
\psfrag{ijk}{$(ijk)$}  
\psfrag{ijl}{$(ijl)$}  
\psfrag{ijm}{$(ijm)$}  
\psfrag{jkl}{$(jkl)$}  
\psfrag{jkm}{$(jkm)$}  
\psfrag{klm}{$(klm)$}  
\psfrag{ikl}{$(ikl)$}  
\psfrag{ikm}{$(ikm)$}  
\psfrag{ilm}{$(ilm)$} 
\psfrag{jlm}{$(jlm)$}
\psfrag{a}{$\alpha$}
\psfrag{b}{$\beta$}
\psfrag{g}{$\gamma$}
\psfrag{d}{$\delta$}
\psfrag{e}{$\epsilon$}
\centering 
\includegraphics[scale=0.8]{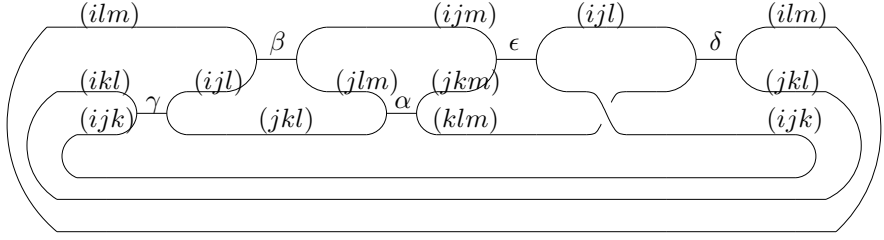} 
\caption{This graph represents the evaluation ${\cal Z}(ijklm)(\alpha \otimes \beta \otimes \gamma \otimes \delta \otimes \epsilon)$ where $(\alpha, \beta , \gamma, \delta, \epsilon)$ is a family of intertwiners.} 
\label{15jsymbol2} 
\end{figure}

We define  the partition function of the 4-simplex 
\begin{eqnarray}
{\cal Z}(ijklm)\;:\; H^+(jklm) \otimes H^+(ijlm) \otimes H^+(ijkl) \otimes  H^-(iklm)  \otimes  H^-(ijkm) \longrightarrow \mathbb C 
\end{eqnarray}
defined by the graph shown in picture (\ref{15jsymbol2}).

Note that the definition of ${\cal Z}(ijklm)$ crucially depends on the ordering of the vertices. However, the sum 
\begin{eqnarray} \label{CYsum}
{\cal Z}_{CY}({\cal T}) = K^{\vert \Delta_0\vert -\vert \Delta_1\vert}
 \sum_{\{j_f\}}
\sum_{\{j_t\}}
 \prod_{f\in \Delta_2} dim_q(j_f) \prod_{t\in \Delta_3} dim_q(j_t)^{-1} 
\prod_{s\in\Delta_4}(15j_q)_s \;,
\end{eqnarray}
does not depend on the chosen ordering nor on ${\cal T}$. It is therefore a topological invariant of the manifold.
This invariant can be expressed solely in term of the signature and the Euler characteristic of the manifold \cite{CKY2}.

 Furthermore, ${\cal Z}_{CY}(M)$ is  the partition function of four dimensional $BF$ theory with cosmological term whose action is:
\begin{eqnarray}
S_{BF,\Lambda} \;=\; \int_M Tr \left(B \; \wedge \; F(A) \; + \; \frac{\Lambda}{2} B\; \wedge \; B \right)\;\;.
\end{eqnarray}
The cosmological constant $\Lambda$ is related to the quantum parameter $q$ by:
\begin{eqnarray}
q=\exp (i \frac{l_P^2}{l_c^2})
\end{eqnarray}
where $l_P = \sqrt{G \hbar}$ is the four dimensional Planck length. This relation is derived in \cite{FK1} for example. 

\subsubsection*{II.2.2. Plebanski Lagrangian with Cosmological Constant}
Pure gravity in $4$ dimensions is a constrained $BF$ theory and the Plebanski lagrangian is precisely designed to implement the contraint through  the use of a Lagrange multiplier \cite{Pl}.

The Einstein-Hilbert action  is written as:
\begin{eqnarray}
S_P[e,\omega] = \frac{1}{G} \int_{ M} d^4 x \; \vert e \vert \; (R_{\mu \nu}^{IJ}[\omega]\;  e_I^{\mu} \; e_J^{\nu}-2\Lambda)
\end{eqnarray}
also  written as: 

\begin{eqnarray}
S_P[e,\omega] = \frac{1}{G} \int_{ M} d^4 x 
\epsilon_{IJKL}( R_{IJ}\wedge e^K\wedge e^L-\frac{ 2\Lambda}{4!}e^I\wedge e^J\wedge e^K \wedge e^L) s(e)
\end{eqnarray}
 where $R=R_{IJ} X^I\wedge X^J$ is the curvature of the connection $\omega$ and  $X^I$ is a   basis of $(\mathbb{R}^4)^{\star}.$ 

When $s(e)=1,$ this is exactly the form of a so(4) $BF$ theory with cosmological constant, with 
$B=B^{IJ}X_I\wedge X_J$ and $B^{IJ}=\epsilon^{IJ}_{KL} e^K\wedge e^L.$ When $B$ is of this form it is said to belong to  the gravitational sector.

Note, as usual,  the sign   $s(e),$ whose possible effect in the quantum theory is still not understood.

Let us add some comments to clarify, for non experts, the condition of simplicity.
If $t$ is an infinitesimal triangle, i.e $t= v\wedge w$ with $v,w$ in the tangent space of a point $x$ in $M,$ 
we can define
 $B_t=<B,v\wedge w>=B^{IJ}_{\mu\nu}X_I\wedge X_J v^\mu w^\nu\in 
(\mathbb{R}^4){}^{\wedge 2}.$
The simplicity condition reads as:
$$B_t\wedge B_t=0,$$ 
which simply means that que quadratic form
 $B_t\wedge B_t$ in $v$ and $w$ vanishes.
The condition of simplicity therefore is equivalent to:
\begin{equation}
B_{(\mu(\nu}^{[IJ} B_{\rho)\sigma)}^{KL]}=0,
\end{equation}
which, when written explicitely in components, is exactly the three conditions: simplicity, intersection and normalisation of \cite{FKP}.

Note, of course, that this condition has nothing to do with the condition $B\wedge B=0$ where the wedge means exterior product simultaneously on forms and on the elements of the Lie algebra $so(4)=({\mathbb R}^4)^{\wedge 2}$, which reads in components
 $$B_{[\mu\nu}^{[IJ} B_{\rho\sigma]}^{KL]}=0, \;\;\text{i.e}\;\;
 \epsilon^{\mu\nu\rho\sigma}\epsilon_{IJKL}
B_{\mu\nu}^{IJ} B_{\rho\sigma}^{KL}=0.$$
When the $B$ field  satisfies the condition that  $B\wedge B$ is nowhere equal to zero,  $B$ is said to be  non degenerate.

It can be shown that the condition of simplicity on $B$ and the condition that $B$ is non degenerate  imply that there exists a cotetrad $e^I$ such that  
\begin{equation}
B^{IJ}=\pm e^I\wedge e^J \:(I\pm)\;\;\text{or }  
B^{IJ}=\pm \epsilon^{IJ}_{KL}e^K\wedge e^L\; (II\pm).
\end{equation}
The gravitational sector is the sector $(II+).$

The Plebanski  Lagrangian is designed to enforce the condition of simplicity on  $B$.

The $SO(\eta)$ Plebanski action is written as:

\begin{equation}
S[\omega,B,\Phi]=
\frac{1}{G}
\int_{M} d^4 x
(F_{IJ}(\omega)\wedge B^{IJ}-
\frac{\Lambda}{4}\epsilon_{IJKL} B^{IJ}\wedge B^{KL}
-\frac{1}{2}\Phi_{IJKL} B^{IJ}\wedge B^{KL}),
\end{equation}

where $\Phi_{IJKL}=\Phi_{[IJ][KL]}$ is symmetric under the exchange of $[IJ]$ with   $[KL]$ and satisfies  the constraint $\epsilon^{IJKL}\Phi_{IJKL}=0.$
An analysis of this theory has been done in \cite{dPF}.

Let us pinpoint the results:
integration on $\Phi$ imposes the constraint that there exists a $4$-form 
 $\lambda$ such that 
 $ B^{IJ}\wedge B^{KL}=\lambda \epsilon^{IJKL}.$
In term of components this is equivalent to 

\begin{equation}
B^{IJ}_{[\mu\nu}B^{KL}_{\rho\sigma]}=
\tilde{\lambda}\epsilon_{\mu\nu\rho\sigma}\epsilon^{IJKL}.\label{simple1}
\end{equation}

When $B$ is non degenerate this condition is also equivalent to 

\begin{equation}
B^{[IJ}_{\mu\nu}B^{KL]}_{\rho\sigma}=
\tilde{\lambda}\epsilon_{\mu\nu\rho\sigma}\epsilon^{IJKL}.\label{simple2}
\end{equation}
Any of these  conditions are equivalent to the condition of simplicity.

As a result, integrating out the Lagrange multiplier $\Phi,$ enforces $B$ to be simple. If one can design a method for restricting even more the $B$ field to   the subsector $(II+),$ one would therefore describe  a quantization of  the gravitational sector with a cosmological constant $\Lambda$.

At the present time we are far from being able to implement and control this restriction.

The  Barrett-Crane model is a spinfoam model where the constraint of simplicity is imposed by a lattice analog of constraint (\ref{simple2}): representations are balanced and the edges are colored by a simple intertwiner, the so-called Barrett-Crane intertwiner.

 The basic idea is to find 
a way to implement the constraints
 (\ref{simple2}) directly in the Ooguri model associated to the group $SO(\eta)$.     

\medskip

The Euclidean Barrett-Crane model is based on simple geometric observations. It consists in describing each $4$-simplex of the triangulation ${\cal T}$ in term of a set of $10$ simple bivectors $b_f \in \Lambda^2 \mathbb R ^4$ associated to its faces $f$. These bivectors satisfy a set of geometrical properties, given in \cite{BC1}. Among them, the simplicity condition, i.e. $b_f \; \wedge \; b_f \;=\;0$, for all bivectors $b_f$, is a discretization of  the 
algebraic conditions (\ref{simple2}).

 Barrett and Crane have postulated that these conditions should be imposed at the  quantum level
 \cite{BC1,BB} as follows.

In the Euclidean case, they first write  $Spin(4)=SU(2)\times SU(2)$ and label its irreducible representations  by couple of non negative half-integers $(j_l,j_r).$
We can decompose this representation with respect to  the diagonal subgroup $SU(2)$ and we denote 
$k=2(j_l-j_r)$ and $i\rho=j_l+j_r+1.$

The Barrett-Crane Euclidean spin foam model  is defined through the following rules:

\begin{enumerate}
\item faces are labelled with finite dimensional irreducible   representations, edges are labelled by intertwiner operators.
\item the representations coloring the faces of the triangulation are simple. Simple representations are of the type $(j,j)$, i.e. $j_l=j_r=j$,  which is equivalent to the fact that $k=0,$ or equivalently that these representations have a non zero vector invariant under the action of the diagonal $SU(2).$
 We denote by ${\cal J}_S$ the set of simple representations. The area of a face $f$ colored by  $(j, j)$ is then given by $a(f)=2l_P^2\sqrt{j(j+1)}$.
\item the intertwiners coloring the tetrahedron of the triangulation is unique and   called Euclidean Barrett-Crane intertwiner. This intertwiner is a 4-valent intertwiner between the simple representations coloring the faces of the tetrahedron and the trivial represenetation. Its  expansion into 3-valent intertwiners introduces only simple representations in the intermediate channel. Up to a normalization, such an intertwiner is uniquely defined \cite{Re2}.
\item
The weight associated to each 4-simplex is constructed from the $10$ simple representations coloring its faces and from the 5 Barrett-Crane intertwiners labelling its boundary tetrahedra. We obtain the so-called $10j$-symbol (see figure \ref{15jsymbol1}) with $i_1, i_2, i_3, i_4, i_5$ fixed to the Barrett-Crane intertwiner.
\end{enumerate}

 Finally, the Euclidean Barrett-Crane partition function is defined by:
\begin{eqnarray} \label{BCsum}
{\cal Z}_{BC}({\cal T}) \; = \; \sum_{\{j_f\}\in {\cal J}_S^{\Delta_2}}
 \prod_{f\in \Delta_2} A_2(j_f)
\prod_{t\in \Delta_3} A_3 (t,\{j_f\})
 \prod_{s\in \Delta_4} A_4(s,\{j_f\})\;\;,
\end{eqnarray}
where $ A_4(s,\{j_f\})$ is the value of the  $10j$ symbol of the colored  simplex $s.$

There is still some confusion about the  value of the  weights $A_2(j_f),
A_3 (t,\{j_f\}).$
An empirical idea, pursued by  \cite{BCHT}, is   to obtain a good balance between all the weights, in such a way that the expectation values of observables do not give trivial results (zero value) or do not  explode to infinity. The work  \cite{BCHT} give a  discussion of 
  this important point and propose weights for $A_2(j_f),
A_3 (t,\{j_f\})$ on a basis of Monte-Carlo simulations. 
   
The goal of  the work \cite{Pe2} is to find in the discretized $BF$ theory the configurations satisfying the constraint of being simple. The final conclusion is that the  configurations of representations and intertwiners which are selected are precisely the one of  the Barrett-Crane  model.
The method, unfortunately,  cannot fix the weights $A_2(j_f),
A_3 (t,\{j_f\}).$

\medskip

The Lorentzian Barrett-Crane model \cite{BC2} is constructed along the same lines, considering instead of $Spin(4),$ the Lorentzian group $SO(3,1)$ or its universal covering $SL(2,\mathbb C)$.

There are two types of models: the first one  includes only spacelike faces and the second  one includes timelike  faces as well.
These models have both been introduced in \cite{BC2}.
A field theory on a group description has been given in \cite{PeRo1,PeRo2}. 

We will only discuss the first model.
 Let us recall that principal irreducible representations of $SL(2,\mathbb C)$ are labelled by a couple $(k,\rho) \in \mathbb Z \times \mathbb R$.

The Lorentzian model of Barrett-Crane is defined using the same recipes as the Euclidean one: the faces are colored by simple representations of type 
$(0,\rho)$ and the tetrahedra are colored by the Lorentzian Barrett-Crane intertwiner.

 The area of a face $f$ is related to the corresponding representation by the formula 
\begin{equation}
a(f) =l_P^2 \sqrt{\rho_f^2+1}.\label{aireface}
\end{equation}

 The Lorentzian Barrett-Crane intertwiner is defined  as in  the Euclidean case: it is unique up to a normalization and can be computed \cite{BC2} using the Haar measure on the classical group (this procedure needs a regularization because of the non-compactness of the group). 

The Lorentzian Barrett-Crane partition function is given by (\ref{BCsum}) where ${\cal J}_S$ is  now the set of  simple representations of $SL(2,\mathbb C)$ of type $(0,\rho).$  and the series is replaced by an integral over $\rho_f.$
Note that the existence of the $10j$ symbol has been proven in \cite{BB2}.

The Lorentzian Barrett-Crane model is conjectured to be an exact  discretization of  the Plebanski action associated to the Lorentz group. 

The sum still suffers from an infrared divergence when $\rho_f$ goes to infinity.
This is why it is very important to add a cut-off on the  $\rho_f$ which can be implemented by adding a cosmological constant and going to the quantization of the Lorentz group. This is the path that we will follow.

Note however that, up to now, another method has been  applied.
 In \cite{PeRo1}, the authors have defined a spin foam model of Lorentzian gravity by choosing appropriate weights $A_2$ and $A_3$. This version  of the Lorentzian model has the marvelous property of being finite in the sense that for each triangulation, the integral over $\rho_f\in \mathbb{R}$ of the state sum is finite \cite{CPR}. Therefore, there is no need to impose a cut off on the $\rho_f$. However the work \cite{BCHT} indicates that this model  may well  be trivial and that other choice of  $A_2$ and $A_3$ are necessary and for them the finiteness seems to be more intricate let alone false.

\medskip

The aim of our work  is to define a quantum deformation of the Lorentzian Barrett-Crane model. Our goal is to introduce a natural cut-off, the cosmological constant, and show that the spin foam model, for a large class of choice of $A_2, A_3$, has the  property of being finite for each triangulation

We will be concentrating on  the case where the quantum parameter is real and related to the cosmological constant by:
\begin{eqnarray}
q=\exp (- \frac{l_P^2}{ l_c^2})\;\;,
\end{eqnarray}
where the cosmological length $l_c$ is defined by $l_c^2=\Lambda^{-1}.$
Though it is consistent with all the results of the litterature, it is still conjectural that Lorentzian Plebanski theory with positive cosmological constant can be described by this type of spin foam model with this regime of $q.$
The computation of the asymptotic of the 4-simplex could  be a useful tool to support this conjecture.

The quantum Lorentzian spin foam model is then the Barrett-Crane model associated to the quantum group $U_q(sl(2,\mathbb C))$ where $q$ is taken to be real. The representation theory of this quantum group is well exposed in 
\cite{BR1} where a Plancherel theorem was proved and the intertwiners between principal unitary representations have been computed in \cite{BR2}. The main results will be recalled in the next section. 

Note  that principal unitary representations of $U_q(sl(2,\mathbb C))$ are labelled by a couple $(k,\rho)$ where $k$ is an integer and $\rho$ is a real  bounded by $\pi\frac{l_c^2}{l_P^2}$. These representations are the quantum analogue of principal representations of $sl(2,\mathbb C)$.
 The quantum Lorentzian model is thus obtained considering only the representations $(0,\rho)$ where $\rho$ is now bounded. The area $a(f)$ of a face colored by $\rho$ is given by (\ref{aireface}), and satisfies therefore the inequalities:
\begin{equation}
l_P^2\leq a(f) \leq \pi l_c^2,
\end{equation}
in the regime where $l_P<<l_c.$

Note that these inequalities are intuitively the expected ones: we cannot define area less than the Planck area and the maximal area of a space like surface  is given by the area of the cosmological horizon.

Before studying the model we recall mathematical results on the quantum Lorentz group and other related objects.

\section*{III. Quantum Hyperboloid and Quantum BC intertwiner.}

In this section, we  introduce the fundamental tools needed to define a cosmological deformation of the  Lorentzian Barrett-Crane model.  We acquaint the reader with the structure of the quantum Lorentz algebra $U_q(sl(2,\mathbb C))$ and its representations. We then  describe the structure of the three dimensional quantum hyperboloid ${\cal H}^{3}_q$ which appears to be a pile of quantum fuzzy spheres. We then define the cosmological deformation of the Barrett-Crane  intertwiner as an integral on ${\cal H}^{3}_q$. The construction of this intertwiner mimics  the classical one \cite{BC2}.  

	\subsection*{III.1. The quantum hyperboloid} 

The classical hyperboloid ${\cal H}^3$ is the  submanifold of 
the four dimensional Minskowski space ${\cal M}^4$ defined  as
\begin{eqnarray}
{\cal H}^3 \; =\; \{x=(x^\mu)_{\mu=0,...,3} \in 
 {\cal M}^4 ,\;  \eta_{\mu\nu}x^\mu x^\nu =-1  \} \;\;,
\end{eqnarray} 
where the metric is given by $\eta_{\mu\nu}=\text{diag}(-1,+1,+1,+1)$.
${\cal H}^3$ admits two connected components ${\cal H}^3_\pm$,  diffeomorphic to $\mathbb R^3$,  which are the two sheets $x^0>0$ and $x^0<0.$

The orthochronous proper Lorentz group $SO(3,1)_+$ acts transitively on  ${\cal H}^3_+$ and therefore ${\cal H}^3_+$ is the orbit of any  element 
$x\in {\cal H}^3_+.$  As a result ${\cal H}^3_+$ is  the  homogeneous space $S_x \backslash  SO(3,1)_+,$ where $S_x$ is the isotropy group (the small group)  of $x.$ This group is  $S_x=SO(3).$
This analysis extends directly to the universal covering group $SL(2,{\mathbb C})$ of  $SO(3,1)_+$, and we get:
 ${\cal H}^3_+ =SU(2)\backslash SL(2,\mathbb{C}).$
The natural measure on Minkowski space induces a measure $dx$ on the 
hyperboloid which is invariant under the action of the Lorentz group.
The same analysis leads to ${\cal H}^3=SO(3)\backslash SO(3,1).$

The homogeneous space $SU(2)\setminus SL(2,\mathbb{C})$ has also two other descriptions.
The first one is associated to   the polar decomposition.
Let \begin{eqnarray}
{\cal M}^4  \rightarrow  \mathbb H \;, \;\; x\mapsto h_x=x^\mu \sigma_\mu= 
\left( \begin{array}{cc} x^0+x^3 & x^1-ix^2 \\ x^1+ix^2 & x^0-x^3\end{array} \right),
\end{eqnarray}   
the isomorphism between  the four dimensional Minkowski space ${\cal M}^4$ and the space $\mathbb H$ of $2\times 2$ Hermitian matrices. 
We denote by $\mathbb H_+$ the subset of positive-definite Hermitian matrices of determinant equal to $1.$
This map identifies the   hyperboloid and  $\mathbb H_+$ :
\begin{eqnarray}
{\cal H}_+^3 \cong {\mathbb H}_+ \; = \; \{h \in {\mathbb H} \; \vert \; \det h = 1 \; \text{and} \; \text{tr}\;h>0 \}.
\end{eqnarray}
The polar  decomposition of the group $SL(2,\mathbb C)$ says that:
\begin{eqnarray} \label{polardecomposition}
\forall g \in SL(2,\mathbb C) \;\;\text{there exists a unique }(k,h) \in SU(2) \times \mathbb H_+\; \text{such that} \; g=kh \;\;.
\end{eqnarray}
As a result the polar decomposition  gives  another  understanding of the  identification  ${\mathbb H}_+ \cong SU(2)\backslash SL(2,\mathbb{C}).$ 
The identification is obtained using the map $
SL(2,\mathbb{C})\rightarrow{\mathbb H}_+, g\mapsto g^\dagger g.$

Another decomposition of $SL(2,\mathbb C)$, the Iwasawa decomposition will provide another description of the hyperboloid.
The Iwasawa decomposition says:
\begin{eqnarray} \label{iwasawa}
\forall g \in SL(2,\mathbb C)
\;\; \text{there exists a unique }(k,a,n) \in SU(2) \times A \times N \; \text{such that} \; g=kan, 
\end{eqnarray}
where $A$ is the group of diagonal positive matrices of determinant 1 and $N$ is the nilpotent group of lower trigonal matrices with diagonal elements equal to 1.
The Iwasawa decomposition implies that the hyperboloid can also be identified as:
\begin{eqnarray}\label{isoH3AN2}
{\cal H}_+^3 \cong\; AN=
 \{  \left( \begin{array}{cc} \lambda & 0 \\ n & \lambda^{-1} \end{array} \right) 
 \; \text{with} \; \lambda \in \mathbb R^{*+} \; \text{and} \; n \in \mathbb C \}.\end{eqnarray} 

Expressing the polar decomposition in term of the Iwasawa one is straightforward and we get:
\begin{eqnarray}
(x_0){}^2  \; =  \; \frac{1}{4}(2+\lambda^2 + \lambda^{-2} +n \overline{n})  && \text{with} \;\; x^0 \geq 1 \;\;, \label{x0C} \\
x_3 \; = \; \frac{1}{4x^0}(\lambda^2-\lambda^{-2}+ n \overline{n}) , &&
x_1+ix_2 \; = \; \frac{1}{2x^0\lambda}n \;\;,
\end{eqnarray}
where $x=(x^\mu)$ is  a point on ${\cal H}_+^3.$ 

The action of the Lorentz group on ${\cal H}_+^3$, is mapped to the natural right action of $SL(2,{\mathbb C})$ on the homogeneous space 
$SU(2) \backslash SL(2,{\mathbb C})$.

Using $x^0$ as a level function, we obtain that ${\cal H}^3_+$ is a cone over $S^2$ foliated by the spheres of equation  $x_1^2+x_2^2+x_3^2=x_0^2-1.$
Note that the radius of the sphere containing the point  $g\in AN,$  is equal to $r^2=\frac{1}{4}(\text{tr }(g^\dagger g)-2).$
These spheres are invariant under the right action of $SU(2)$ on $SU(2) \backslash SL(2,{\mathbb C}).$

The measure on the hyperboloid is identified to the measure on $SU(2) \backslash SL(2,{\mathbb C})$ invariant under the right action of $SL(2,{\mathbb C})$.
As a result if we identify $SU(2) \backslash SL(2,{\mathbb C})$ with $AN$, the Haar measure on $SL(2,{\mathbb C})$ factorizes as $dg=dk\; d(an)$ where $d(an)$ is the right invariant  integral on $AN$. The measure on the hyperboloid is therefore identified with $d(an)$.

We will now describe a construction of  the quantum algebra $U_q(an)$ and of its algebra of function $F_q(AN).$ This is obtained through the use of the quantum duality principle.
After having completed this task, we introduce the notion of quantum hyperboloid.

The quantum Lorentz algebra $U_q(sl(2,\mathbb C))$ and   the algebra $SL_q(2,\mathbb C)$ of functions on this quantum group is recalled briefly in the appendix A.1. The quantum Lorentz algebra  is  the quantum double of $U_q(su(2))$. This construction is in this case the quantum analog of   the  Iwazawa decomposition of the Lorentz algebra i.e. $sl(2,\mathbb C) \; = \; su(2) \; \oplus \; an(2)$ where $an(2)$ is the real algebra of $2\times 2$ lower triangular complex matrices with real diagonal of zero trace.
 A complete survey of this construction is   exposed in \cite{BR1}.

\medskip

We let $q=e^{-\kappa} \in \rbrack 0, 1 \lbrack$ and define  $A=U_q(su(2))$, to be the Hopf algebra generated  by the elements  $J_{\pm}$, $q^{\pm J_z}$, and the relations: 

\begin{eqnarray}  
q^{\pm J_z}  q^{\mp J_z} \; = \; 1\;,\;\;\;\;  q^{J_z} J_{\pm} q^{-J_z} \; = \; q^{\pm 1} J_{\pm} \; , \;\;\;\; \lbrack J_+,\, J_- \rbrack \; = \; \frac{q^{2 J_z} \; - \; q^{-2J_z}}{q \; - \; q^{-1}} \;\;\; .  \end{eqnarray}  

This  algebra can be endowed with a structure of  ribbon quasi-triangular Hopf algebra. We will only recall the structure of coproduct which reads:

$$\Delta(q^{\pm J_z})=q^{\pm J_z}\otimes q^{\pm J_z}, 
\Delta(J^{\pm})=q^{-J_z}\otimes J_{\pm}+J_{\pm}\otimes q^{J_z}.$$ 
We will denote by $R$ or $R^{(+)}$ the $R$-matrix and by  $R^{(-)}=R_{21}^{-1}$.
 
 Finite dimensional unitary irreducible representations $\stackrel{I}{\pi}$ of $U_q(su(2))$ are labelled by a spin $I$ and let us define $\stackrel{I}{V}$  the vector space associated to the representation $\stackrel{I}{\pi}$. 
This  representation is of dimension $2I+1$ and we denote by
 $\{\stackrel{I}{e}_m , m=-I,\cdots,I\}$ the orthonormal basis  such that the action of the generators on this basis is given by the following expressions:
 \begin{eqnarray}  \label{SU2rep}
q^{J_z} \; \stackrel{I}{e}_m \; & = & \; q^{m} \; \stackrel{I}{e}_m \; , \\  
J_{\pm} \; \stackrel{I}{e}_m \; & = & \; q^{\pm \frac{1}{2}} \sqrt{\lbrack I \pm m +1 \rbrack_q \lbrack I \mp m \rbrack_q }\stackrel{I}{e}_{m \pm 1}\;\; . \nonumber 
\end{eqnarray}
where we have denoted as usual $[z]_q=\frac{q^z-q^{-z}}{q-q^{-1}}.$
The quantum dimension of this representation is defined by $ Tr(\stackrel{I}{\pi}(\mu))=[2I+1]_q$ where $\mu=q^{2J_z}.$
We denote by 
$\{\stackrel{I}{e}{}\!^m \vert m=-I,\cdots,I\}$ the  dual basis.

 Through the use of the coproduct,  representations can be tensorized. For any unitary representations $\stackrel{I}{\pi},\stackrel{J}{\pi},\stackrel{K}{\pi}$, we define the Clebsch-Gordan maps $\Psi^{K}_{IJ}$ (resp. $\Phi^{IJ}_{K}$) as intertwiners from $\stackrel{I}{V}\otimes\stackrel{J}{V}$ to $\stackrel{K}{V}$
 (resp. intertwiners from
 $\stackrel{K}{V}$ to $\stackrel{I}{V}\otimes\stackrel{J}{V})$. 
With a particular choice of normalisation (see \cite{BR1}), these elements are unique and their matrix elements are  the 3j coefficients of  $U_q(su(2)).$   

We will denote by   $\stackrel{I}{k}{}\!\!^j_i$ the matrix elements of $ \stackrel{I}{\pi}$, i.e the linear forms on $A,$ $\stackrel{I}{k}{}\!\!^j_i=<\stackrel{I}{e}{}\!^j \vert \stackrel{I}{\pi}(\cdot) \vert \stackrel{I}{e}_i >.$
The algebra of quantum deformations of polynomial functions on $SU(2)$ is the Hopf-algebra generated as a vector space by   $\stackrel{I}{k}{}\!\!^j_i$ and denoted $F_q(SU(2)).$

By a direct application of the definition, the algebra structure is given by:
\begin{eqnarray}
\stackrel{I}{k}_1 \stackrel{J}{k}_2 = \sum_K \Phi^{IJ}_M \stackrel{M}{k} \Psi^M_{IJ}
\end{eqnarray}
where we have denoted, as usual, $\stackrel{I}{k}_1 = \stackrel{I}{k} \otimes 1$, $\stackrel{I}{k}_2 = 1 \otimes \stackrel{I}{k}$ and $\stackrel{IJ}{R} = (\stackrel{I}{\pi} \otimes \stackrel{J}{\pi})(R)$.

The existence of the $R$ matrix implies the braiding relations:
\begin{equation}\label{PolSU2} \stackrel{IJ}{R}_{12} \stackrel{I}{k}_1 \stackrel{J}{k}_2 \; = \; \stackrel{J}{k}_2 \stackrel{I}{k}_1  \stackrel{IJ}{R}_{12}\;.
\end{equation}
Because the spin $\onehalf$ when tensored sufficiently many times contains any prescribed spin, 
$F_q(SU(2))$ can equivalently be viewed as the Hopf algebra generated by the matrix elements of $\stackrel{\frac{1}{2}}{k}$ with the defining relation (\ref{PolSU2}) and the  additional relation $det_q(\stackrel{\frac{1}{2}}{k})=1.$
Explicit relations are recalled in the appendix of \cite{BR1} for example.

A very important notion in the theory of quantum groups is the principle of quantum duality.
We will only give the general idea of this principle, leaving aside delicate issues related to the topology of the spaces.
Let  ${\mathfrak g}$  a Lie algebra. If $U_q({\mathfrak g})$ is a quantization, of Hopf algebra type,  of the enveloping algebra  of ${\mathfrak g}$, it endows a structure of Lie algebra on the dual of the Lie algebra  ${\mathfrak g}^*$. It is obtained as follows: if 
$a\in {\mathfrak g}$, we denote $q=1-\kappa+\sum_{n\geq 2}a_n \kappa^{n},$ we can define $\delta(a)=
\lim_{\kappa\rightarrow 0}\frac{\Delta(a)-\Delta_{21}(a)}{\kappa}.$ It can be shown that $\delta:{\mathfrak g}\rightarrow {\mathfrak g}\wedge {\mathfrak g}.$ As a result the transpose map $\delta^t$ maps $ {\mathfrak g}^*\times  {\mathfrak g}^*\rightarrow {\mathfrak g}^*$, is antilinear and satisfies Jacobi identity because $\Delta$ is coassociative.
The principle of ``quantum duality'' expresses the fact that:
if ${\mathfrak g}$  is a Lie algebra, such that $U_q({\mathfrak g})$ is a quantization of $U({\mathfrak g})$, then 
 ${\mathfrak g}^*$ admits a quantization, denoted,  $U_q({\mathfrak g}^*)$, and we have $U_q({\mathfrak g}^*)=(U_q({\mathfrak g}))^*$ as Hopf algebras.

A typical example is ${\mathfrak g}=su(2).$ In this case a direct computation, using the structure of $U_q(su(2))$ shows that $su(2)^*=an(2).$ Therefore we can define, through the use of the quantum duality principle:
 $U_q(an(2))=(U_q(su(2)))^*.$ 

Note that such definition can only hold at the non-commutative level i.e when $q$ is non equal to $1.$ Indeed when $q=1$, $U_q(su(2))$ is cocommutative, 
therefore the dual $U_q(su(2))^*$ is a commutative algebra, which prevents to be isomorphic to $U_q(an(2))$
This principle can also be used at the level of function spaces. Let $G$ be the
 Lie group of Lie algebra ${\mathfrak g}$, let $G^*$ be the Lie group of Lie algebra 
 ${\mathfrak g}^*.$ We can define $F_q(G)= (U_q({\mathfrak g}))^*, 
F_q(G^*)= U_q({\mathfrak g^*})^*$. The quantum duality principle implies that we have 
$F_q(G)^*=F_q(G^*)$ as Hopf algebras.

As a result, in our case, $G=SU(2)$, $G^*=AN,$ the quantum duality principle tells us that we can define $F_q(AN)=(F_q(SU(2)))^*.$

We will now describe the space $F_q(AN)$, which is a star-Hopf algebra. We will also show that it is very natural to define the quantum hyperboloid $H^3_{+q}$ as being, as in the classical case, the Hopf algebra  $H_{+q}^3=F_q(AN)$.

We have already described the structure of Hopf algebra of $F_q(SU(2)).$

We can now define spaces of functions of different classes on $AN$. The first one is the quantization of the  compact supported functions on  $AN,$ it is defined as:
\begin{equation}
Fun_c(AN_q)=\bigoplus_{I\in\frac{1}{2}{\mathbb N}}Mat_{2I+1}({\mathbb C})
\end{equation}
with the following (``multiplier'')-Hopf algebra structure:
\begin{eqnarray}
(\oplus_I a_I)(\oplus_I b_I)=\oplus_I a_I b_I\label{multfunc}\\ 
\Delta(\oplus_I a_I)=\sum_{I,J,K} \Phi^{JK}_I a_I \Psi_{JK}^I\label{coprodfunc}\\ 
(\oplus_I a_I)^\star=\oplus_I a_I^\dagger,\label{starfunc}
\end{eqnarray}
which is the dual of the Hopf algebra $F_q(SU(2)).$
Note that by definition an element $(\oplus_I a_I)$ in $Fun_c(AN_q)$ is such that only a finite number of $a_I$ are non zero.

The coproduct maps $Fun_c(AN_q)$ into the larger space 
$\prod_{I}Mat_{2I+1}({\mathbb C})\otimes \prod_{I}Mat_{2I+1}({\mathbb C})$ which is the reason why we have to work in the scheme of multiplier Hopf algebras.

The quantum algebra of functions (generally unbounded) on $AN$ is denoted $Fun(AN_q)$ and defined as 
\begin{equation}Fun(AN_q)=\prod_I Mat_{2I+1}({\mathbb C}).
\end{equation}
It can be endowed with a structure of Hopf algebra with the same formulas for the product, the coproduct and the star as (\ref{multfunc}, \ref{coprodfunc},
\ref{starfunc}) .
It also possess a unit element defined by $1=\prod_{I}1_{I}$ where $1_I$ is the identity matrix of $Mat_{2I+1}({\mathbb C})$.

The very definition of  $Fun_c(AN)$ comes from the existence of a Haar measure.
$Fun_c(AN_q)$ admits a right invariant integral 
$h_{AN}: Fun_c(AN_q)\rightarrow {\mathbb C}$ defined as follows:
\begin{equation}
h_{AN}(\oplus_I a_I)=\sum_I \text {tr}(a_I \stackrel{I}{\pi}(\mu)),
\end{equation}
which satisfies $(id\otimes h_{AN})\Delta(x)=h_{AN}(x) 1$ for any $x\in Fun_c(AN_q).$

We  define  $Fun_c(H^3_{+q})$, the  algebra of  compact supported functions on the quantum hyperboloid,  as   being, as in the classical case, the  algebra  $Fun_c(H^3_{+q})=Fun_c(AN_q)$.

Therefore, as an algebra, it has the structure
 $Fun_c(H^3_{+q})=\oplus_{I}Mat_{2I+1}({\mathbb C}).$

This description is the deformation of the foliation of $H^3_{+}$ by quantum fuzzy spheres. Quantum fuzzy spheres have been introduced and studied in 
\cite{GMS}. The construction of the quantum hyperboloid that we are proposing has a scaled limit $q\rightarrow 1$    studied recently by \cite{BM}.
We will see more on this topic in the sequel.

We shall now group together the quantization of $su(2)$ and $an$ in order to obtain $U_q(sl(2,{ \mathbb C}))$. This is done through the construction of the quantum double exposed in the appendix A.1.

\medskip

Let us now introduce the quantum algebra  of functions on
 $SL(2,{\mathbb C})$,  good references on this subject being \cite{PW,BR1}.

We define the space of compact supported functions on $SL(2,{\mathbb C})$ by
$Fun_c(SL(2,{\mathbb C})_q)=Fun(SU_q(2))\otimes Fun_c(AN_q)$ as algebras.
The expression of the coproduct is written in \cite{BR1}.
There exists a Haar measure $h$ on 
$Fun_c(SL(2,{\mathbb C})_q)$ which  expression is 
$h=h_{SU(2)}\otimes h_{AN}$, where $h_{SU(2)}$ denotes the normalized Haar measure on $Fun(SU_q(2))$. The functional $h$ is a right and left integral for $Fun_c(SL(2,{\mathbb C})_q).$

The matrix elements of the 2-dimensional left spinorial representation of 
$U_q(sl(2,\mathbb C))$ can be seen as quantum deformations of polynomial functions on 
$SL(2,{\mathbb C}).$ We denote by $G$ this $2\times 2$ matrix of non commuting elements.
In \cite{PW,BR1} the Iwasawa decomposition of this matrix has been done, and we have $G=UT$ where  $U$ is a   $2\times 2$ matrix satisfying the defining relations of $SU_q(2)$ and $T$ is a lower triangular matrix of the form:

\begin{equation}
T=   \left( \begin{array}{cc}\hat{\lambda} & 0 \\ (\frac{q^2+1}{2})^{1/2}\hat{n} & \hat{\lambda}^{-1} \end{array} \right) 
 \;
\end{equation}
with the relations:

 \begin{equation}
{\hat \lambda}{\hat \lambda}^{\star}={\hat \lambda}^{\star}{\hat \lambda},\;\;\; {\hat \lambda}{\hat  n}=q^{-1}{\hat n} {\hat \lambda}, \;\;
{\hat \lambda}{\hat n}^{\star}=q {\hat n}^{\star} {\hat \lambda},\;\;\;
  {\hat n} {\hat n}^{\star}-{\hat n}^{\star}
{\hat  n}=\frac{2}{q+q^{-1}}(q^{-1}-q)({\hat \lambda}^2-{\hat \lambda}^{-2}).
 \end{equation}
Note that these relations, which can be seen as the defining relations of 
$F_q(AN)$ are also the defining relations of $U_q(su(2))$ (this is the quantum duality principle) through the identification:

\begin{eqnarray} 
\hat{\lambda} =q^{J_z},\;\;
 \hat{n} = q^{-\frac{1}{2}}(q-q^{-1})(\frac{2}{q+q^{-1}})^{\frac{1}{2}}J_-, \;\; \hat{n}^* = q^{+\frac{1}{2}}(q-q^{-1})(\frac{2}{q+q^{-1}})^{\frac{1}{2}}J_+.
\end{eqnarray}

Therefore the element ${\hat \lambda}$ is  represented in $Fun_c(AN_q)$ as being the elements ${\hat \lambda}=\prod_I\stackrel{I}{\pi}(q^{J_z})$ with the analog   representation for ${\hat n}$ and $ {\hat n}^*.$

It is explained in \cite{BR1} why the classical limit of the  integral
 $h_{AN}$ on $Fun_c(AN_q)$ is the continuous integral 
$\lambda^{-3}d\lambda dn d{\bar n}.$ 

Of particular importance for the understanding of this classical limit is the 
 element $\hat{C}=\frac{1}{2}(\hat{n} \hat{n}^* + \hat{n}^* \hat{n}) +(\hat{\lambda}^2 + \hat{\lambda}^{-2}).$ Note that the classical limit of this element is exactly $C=4r^2+2$ where $r$ is the radius of the sphere.
In the quantum case, $\hat{C}$ is in the center of the algebra and therefore is represented by a scalar $C_I$ on each subspace of matrices $Mat_{2I+1}({\mathbb C})$. Its value is 
$$C_I=2\frac{q^{2I+1}+q^{-2I-1}}{q+q^{-1}}.$$
The value of the radius of the fuzzy sphere $Mat_{2I+1}({\mathbb C})$ is therefore defined by $C_I=4r_I^2+2$ and explicitely given by $2(q+q^{-1})r_I^2=(q-q^{-1})^2[I][I+1].$

The quantum analog of the relation (\ref{x0C}) is $\hat{X}_0^2 \; = \; \frac{1}{4}(2+\hat{C})$. Since $\hat{C}$ is a Casimir with positive eigenvalues, the time component $\hat{X}_0$  is a central element, with discrete spectrum 
\begin{eqnarray}
\hat{X}_0 \; = \; \sqrt{\frac{1}{2} \left( 1 + \frac{\cosh (2I+1) \kappa}{\cosh  \kappa} \right)}
\end{eqnarray}
on each components of $Mat_{2I+1}({\mathbb C})$.  

The classical limit, $\kappa\rightarrow 0,$ is recovered as follows.
A quantum fuzzy sphere indexed by $I$ recovers its classical limit when 
  $(2I+1)\kappa$ tends to a classical value $l.$ 
In that case, this classical sphere is located at time 
$X_0= \cosh \frac{l}{2}.$ The value 
$l/2$ is interpreted as the distance on the hyperboloid between any point on this classical sphere and the point located at   $X_0=1.$

	\subsection*{III.3. Simple representations and the q-BC  Lorentzian intertwiner}
\medskip

Let us now recall basic notions on the representation theory of  the quantum Lorentz group. Irreducible unitary representations of $U_q(sl(2,\mathbb C))$ have been classified. They include the principal series on which the Plancherel
 measure is concentrated. The representations of the principal series are 
 labelled by a couple $\alpha=(\alpha^l,\alpha^r)\in \mathbb{C}^2$ with 
$2(\alpha_l-\alpha_r)=k, \alpha_l+\alpha_r+1=i\rho$ with 
$k\in{\mathbb Z},\rho \in ]-\frac{\pi}{\kappa}, \frac{\pi}{\kappa}[$. We denote $\stackrel{\alpha}{\Pi}$ the principal representation associated to $\alpha$ acting on the vector space  $\stackrel{\alpha}{{\mathbb V}}.$  As explained  in \cite{BR1},
 $\stackrel{\alpha}{\Pi}$   is a Harisch-Chandra representation  which decomposes  in representations of $U_q(su(2))$   as:
\begin{eqnarray} \label{module}
\stackrel{\alpha}{{\mathbb V}} \; = \; \bigoplus_{I,I-k \in \mathbb N} \stackrel{I}{V} \;\;.
\end{eqnarray}
A basis of  $\stackrel{\alpha}{{\mathbb V}}$ is  given by $(\stackrel{I}{e}_m(\alpha))_{I, I-k \in \mathbb N}$  where, for $I$ fixed, $(\stackrel{I}{e}_m(\alpha))_{m}$ is the basis of the $U_q(su(2))$ module $\stackrel{I}{V}$ (\ref{SU2rep}).   $\stackrel{\alpha}{{\mathbb V}}$ contains a $U_q(su(2))$-invariant vector if and only if $k=0$. 
These representations are called 
simple representations. If $\alpha$ is simple, we will denote $\omega(\alpha)$ the  $U_q(su(2))$ invariant vector of unit length $\omega(\alpha)= \stackrel{0}{e}_0(\alpha).$
As a result the matrix element 
$K_\rho=<\omega(\alpha)\vert  \stackrel{\alpha}{\Pi}(.)\vert\omega(\alpha)> $ is a linear form on
 $U_q(sl(2,{\mathbb C}))$. We will also use the notation 
$\stackrel{\alpha}{K}=K_\rho$. $ K_\rho$ is an element of
 $F_q(SL(2,{\mathbb C})$. 
Because this function is invariant on the left by $U_q(su(2))$, the Iwasawa decomposition implies that $K_\rho\in F_q(AN).$ As a result $K_\rho$ defines a function on the quantum hyperboloid.
Moreover this function is invariant under the right action of $U_q(su(2)).$ As a consequence it is diagonal: $K_\rho=\sum_J K_\rho(J) 1_J$ with $ K_\rho(J)\in {\mathbb C}.$

We can compute exactly its expression. This is done using the work \cite{BR1} and the necessary 
background is recalled in appendix A.1: we have $ K_\rho(J)=\Lambda^{JJ}_{00}(\alpha).$
We finally obtain 
\begin{equation} \label{quantumkernel}
K_\rho=\sum_J\frac{[i(2J+1)\rho]}{[i\rho][2J+1]}1_J.
\end{equation}
Note that $K_\rho$ satisfies $S(K_\rho)=K_\rho.$

It is an easy check to see that in the classical limit, i.e $\kappa\rightarrow 0$,
 $(2J+1)\kappa\rightarrow l,$ the coefficient 
\begin{equation} K_\rho(J) \rightarrow \frac{\sin (l\rho)}{\rho \sinh (l)},
\end{equation} 
which is exactly the classical zonal function  used in \cite{BC2}.

\medskip

The Barrett-Crane model is defined through the use of a particular intertwiner, which is nowadays called Barrett-Crane intertwiner (BC intertwiner). 
Its quantum deformation is easy to define and this is the purpose of what is coming next.

It will be very important in the sequel to generalize to the quantum 
deformation the following type of integrals.
If $\Pi$ is a  principal representation of  
$sl(2,{\mathbb C}),$ we also denote $\Pi$ the associated unitary representation of the group $SL(2,{\mathbb C}).$
If $\varphi$ is a smooth compact supported function on $SL(2,{\mathbb C})$, 
the following bounded operator can be defined:
\begin{equation}
\Pi(\varphi)=\int \Pi(g)\varphi(g)dg.
\end{equation} 
These type of integrals have a direct generalization to the quantum group case although there is no more notion of point nor of the meaning of
 $\Pi(g).$
Let $\Pi$ be a principal representation of $U_q(sl(2,{\mathbb C}))$, 
we denote $(x_A)$ the  basis of $Fun_q(SL(2,{\mathbb C}))$ as defined in the 
appendix A.1. $(x^A)$  the dual basis of  $(x_A),$ is a basis of  
 $U_q(sl(2,{\mathbb C})).$ As a result we can define $\Pi(x^A),$ its exact expression in a basis being given in the Appendix A.1.
We have shown in \cite{BR1} (eq. 93) that if $ \varphi$ is an element of 
$Fun_c(SL_q(2,\mathbb C))$ the following operator is well defined and is moreover of finite rank and corank (i.e it is a finite dimensional matrix):
\begin{equation}
\sum_A \Pi(x^A)h(x_A\varphi).
\end{equation}
This  operator is the quantum analog of $\Pi(\varphi).$
More generally if $\{\stackrel{\alpha_i}{\Pi}, i=1,...,n\}$ is a family of principal representations of 
 $SL(2,{\mathbb C})$, integrals of the type 
\begin{equation}
(\bigotimes\limits_{i=1}^n\stackrel{\alpha_i}{\Pi})(\varphi)=
\int \bigotimes\limits_{i=1}^n \stackrel{\alpha_i}{\Pi}(g) \varphi(g)dg,
\end{equation}
often appear in the construction of interwiners.

They will be generalized as follows:
\begin{eqnarray}
(\bigotimes\limits_{i=1}^n\stackrel{\alpha_i}{\Pi})(\varphi)=
 (\bigotimes\limits_{i=1}^n \stackrel{\alpha_i}{\Pi})\Delta^{(n)}(x^A)h(x_A \varphi)=\\ \nonumber
=\sum_{A_1,...,A_n} \bigotimes\limits_{i=1}^n \stackrel{\alpha_i}{\Pi}(x^{A_i})h(\prod_{i=1}^n x_{A_i} \varphi)
\end{eqnarray}

We will have to extend  these operators in the case where $\varphi$ is the constant function $1$. In particular we will show that when $n\geq 3$, the operator 
$(\bigotimes\limits_{i=1}\stackrel{\alpha_i}{\Pi})(1)$ is well 
defined. Note that this operator, in the case $n=3,$ has already been studied in detail in \cite{BR2}. 

The basic definition of the  q-deformation of the Lorentzian BC  intertwiner is similar to the classical one. The q-BC Lorentzian intertwiner is a n-valent intertwiner between simple representations of the quantum Lorentz group such that  any expansion of it into 3-valent intertwiners introduces only simple representations in the intermediate channel.

\begin{figure} 
\psfrag{a1}{$\alpha_1$}  
\psfrag{a2}{$\alpha_2$}  
\psfrag{a3}{$\alpha_3$}  
\psfrag{a4}{$\alpha_4$} 
\psfrag{an}{$\alpha_n$} 
\psfrag{b1}{$\beta_1$}  
\psfrag{b2}{$\beta_2$}  
\psfrag{bp}{$\beta_p$}
\centering 
\includegraphics[scale=0.8]{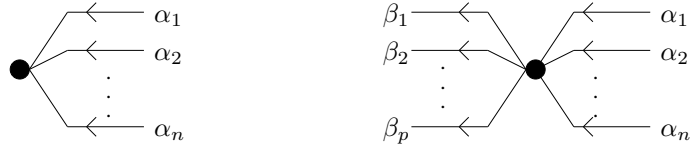} 
\caption{The q-BC Lorentzian intertwiner and its generalization.}  
\label{Intertwiner} 
\end{figure}

\begin{definition}{q-BC Lorentzian intertwiner.} 

Let $\alpha=(\alpha_1,\cdots,\alpha_n)$ a sequence  of simple representations of the quantum Lorentz group. We will denote
 $\mathbb{V}[\alpha]=\bigotimes\limits_{i=1}^{n}
\stackrel{\alpha_i}{\mathbb V}$ and
 $\omega[\alpha]= \bigotimes\limits_{i=1}^{n}\omega(\alpha_i).$
The Quantum Lorentzian intertwiner
 $\iota[\alpha]:\mathbb{V}[\alpha]
\rightarrow {\mathbb C}$ is defined by:
\begin{eqnarray} \label{QLI}
\iota[\alpha] \; = 
<\omega[\alpha]\vert
(\bigotimes\limits_{i=1}^n\stackrel{\alpha_i}{\Pi})(1).
\end{eqnarray}
 It is represented by the left graph of caption (\ref{Intertwiner}). 
\end{definition}

From its formal definition, $\iota[\alpha]$, if it exists, is an interwiner.
We are therefore led to study the convergence property of the series defining $\iota[\alpha].$
We note first that, for $z\in \mathbb{V}[\alpha]$,  
$\iota[\alpha](z)=h(f[\alpha](z))$ where $f[\alpha](z)$  is an element of 
$Fun(AN_q)$ because the vector $\omega[\alpha]$  is invariant  by the right  action of $U_q(su(2)).$ 
 As a consequence, the q-BC  Lorentzian intertwiner is defined by an integral on the quantum hyperboloid whose convergence  will be analyzed in the sequel.

{}From the expression of the matrix element $\stackrel{\alpha}{\Pi}$ (see the appendix A.1), it is straigthforward to see that:
\begin{eqnarray}<\omega(\alpha) \; \vert \; \stackrel{\alpha}{\Pi} \; \vert \; \stackrel{A}{e}_a\!\!(\alpha)> \; = \; \sum_M  \Clebpsi{M}{A}{M}{m'}{a}{m} \Lambda^{MM}_{0A}(\alpha) \; \stackrel{M}{E}{}\!^{m'}_m.
\end{eqnarray}
As a result we obtain:
\begin{proposition}
Let $\alpha = (\alpha_1,\cdots, \alpha_n)$ a family of simple representations, 
let $A=(A_1,\cdots,A_n)$ a family of spins and $a=(a_1,\cdots,a_n)$ the  magnetic numbers. We denote by $\stackrel{A}{e}_{a}[\alpha] = \bigotimes_{i=1}^n  \stackrel{A_i}{e}_{a_i}\!\!(\alpha_i)$  and we have:
\begin{eqnarray} \label{sum}
\iota[\alpha]\;\vert \; \stackrel{A}{e}_{a}[\alpha]> \; = \; \sum_M [d_M] \; C(M,A,a) \; \prod_{i=1}^n \Lambda_{0A_i}^{MM}(\alpha_i),
\end{eqnarray}

where
\begin{eqnarray}
C(M,A,a) \; = \; \sum_{m_1,...,m_{n+1}} (\stackrel{M}{\mu}{}\!^{-1})^{m_1}_{m_{n+1}}  
\prod_{p=1}^n \Clebpsi{M}{A_p}{M}{m_p}{a_p}{m_{p+1}} \;\;.
\end{eqnarray}
The series (\ref{sum}) converges when $n\geq 3$ and in this case the q-BC  Lorentzian intertwiner (\ref{QLI}) is well defined.
\end{proposition}
\Proof \\
The Clebsch-Gordan coefficients are bounded by one, as a result we get the bound:
\begin{equation}
\vert C(M,a,A)\vert \leq [d_M](2M+1)^n.
\end{equation}
{}From the asymptotics computed in \cite{BR2}, we have the following bound  when $M$ goes to infinity:
\begin{equation} \label{boundofLambda}
 \Lambda_{0A_i}^{MM}(\alpha_i)={\cal O}(M q^{2M}).
\end{equation}
As a consequence, 
$ [d_M] \; C(M,A,a) \; \prod_{p=1}^n \Lambda_{0A_p}^{MM}(\alpha_p) 
 ={\cal O}(M^{2n} q^{2M(n-2)})$ which implies that the  series (\ref{sum}) converges
 absolutely when $n\geq 3.$  $\Box$

\medskip

We will now prove that the four valent q-BC intertwiner satisfies the property that 
its expansion  in three valent intertwiners only includes simple representations in the
intermediate channel:

\begin{proposition} \label{decompositionoftheqBCintertwiner}
The four valent q-BC Lorentzian intertwiner $\iota[\alpha]$ is  decomposed in three valent $U_q(sl(2,\mathbb C))$ intertwiners as follows:
\begin{eqnarray} \label{decomposition}
\iota[\alpha] \; = \; \int d\beta P(\beta)\; F(\alpha,\beta) \; \Psi_{\beta,\beta}^{\;\;0}
( \Psi_{\alpha_1,\alpha_2}^{\;\;\beta} \otimes \Psi_{\alpha_3,\alpha_4}^{\;\;\beta}) 
\end{eqnarray}
where the measure $P(\beta) d\beta $  is  the Plancherel measure and the  
 kernel $F(\alpha,\beta)$ has its support  on the set of simple 
 representations. Its explicit expression is given in the following proof. 
\end{proposition}
\Proof \\
The proof of the decomposition (\ref{decomposition}) is a direct application 
of the  relations (\ref{orthogonalityrelation},\ref{SchurLemma}) given in the appendix.

Indeed, one first applies the completeness relation (\ref{orthogonalityrelation}) to the q-BC intertwiner:
\begin{eqnarray*}
\iota[\alpha] \; = \; <\omega[\alpha] \vert
 \int d\beta  d\gamma \;
M(\alpha_1,\alpha_2,\beta)
M(\alpha_3,\alpha_4,\gamma)
 \Phi^{\alpha_1  \alpha_2}_{\;\; \beta} \otimes \Phi^{\alpha_3  \alpha_4}_{\;\; \gamma} 
(\stackrel{\beta}{\Pi} \otimes \stackrel{\gamma}{\Pi})(1)
 \Psi^{\;\; \beta}_{\alpha_1 \alpha_2} \otimes \Psi^{\;\; \gamma}_{\alpha_3 \alpha_4}\;\;.
\end{eqnarray*}
The use of the orthogonality relation  (\ref{SchurLemma}) implies:
\begin{eqnarray*}
\iota[\alpha] \; = \; < \omega[\alpha] \vert \int d\beta \; \frac{1}{P(\beta)}M(\alpha_1,\alpha_2,\beta)
M(\alpha_3,\alpha_4,\beta)
 \Phi^{\alpha_1  \alpha_2}_{\;\; \beta} \otimes \Phi^{\alpha_3  \alpha_4}_{\;\; \beta}
\Phi^{\beta,\beta}_{\;\;0}  \; \Psi_{\beta,\beta}^{\;\;0} \; \Psi^{\;\; \beta}_{\alpha_1 \alpha_2} \otimes \Psi^{\;\; \beta}_{\alpha_3 \alpha_4}\;\;.
\end{eqnarray*}

Moreover the Clebsh-Gordan coefficient $<\omega(\alpha_1)\otimes \omega(\alpha_2) \vert \Phi^{\alpha_1 \alpha_2}_{\;\;\beta} \vert v>$ with $v \in \; \stackrel{\beta}{\mathbb{V}}$ is different from zero if and only if $v$ is a non zero  $U_q(su(2))$-invariant vector. Such a vector exists if $\beta$ is a simple representation and by definition the normalization of $\Phi^{\alpha_1 \alpha_2}_{\;\;\beta}$ is chosen such that  $<\omega(\alpha_1) \otimes \omega(\alpha_2) \vert \Phi^{\alpha_1 \alpha_2}_{\;\; \beta} \vert \omega(\beta)> =1.$
As a consequence, the q-BC intertwiner can be decomposed as (\ref{decomposition}) where  
\begin{equation}
F(\alpha,\beta)=
\frac{1}{P(\beta)^2}M(\alpha_1,\alpha_2,\beta)M(\alpha_3,\alpha_4,\beta).
\end{equation}
 $\Box$

Note that we have proven the property of decomposition of the q-BC Lorentzian
 intertwiner in  a particular channel.
It is straightforward to extend this proof to the two other channels, as in
  the classical case \cite{Re2}.

A very important property satisfied by the q-BC intertwiner is the invariance under braiding, 
which can be formulated as follows.

Let $\alpha=(\alpha_1,...,\alpha_n)$ be an n-uplet of principal representations and let $b$ be an element of the braid group on $n$ strands. To $b$ is associated a unique permutation on $n$ points, denoted $\sigma(b).$ We associate, as usual, to  $b$ a linear operator ${\mathbb R}(b): 
\mathbb{ V}[\alpha] \rightarrow \mathbb{ V}[\sigma(b)(\alpha)]$ by representing the braid by a product of ${\mathbb R}$ matrices of $U_q(sl(2,{\mathbb C}))$.
Note that this is well defined because the ${\mathbb R}$ matrix of $U_q(sl(2,\mathbb{C}))$ evaluated in the tensor product of principal representation is well defined, which is a non trivial fact (see \cite{BR2}).

\begin{figure} 
\psfrag{a1}{$\alpha_1$}  
\psfrag{a2}{$\alpha_2$}  
\psfrag{a3}{$\alpha_3$}  
\psfrag{a4}{$\alpha_4$} 
\psfrag{=}{$=$} 
\centering 
\includegraphics[scale=0.8]{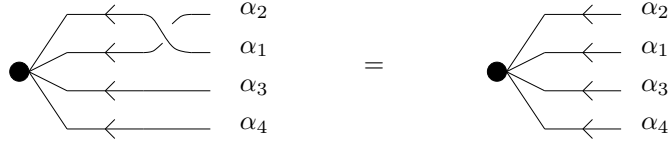} 
\caption{Invariance of the Quantum Lorentzian intertwiner by braidings.}  
\label{Braiding} 
\end{figure}

\begin{proposition}
The property of invariance under braiding states that:
\begin{equation}
\iota[\sigma(b)(\alpha)]{\mathbb R}(b)=\iota[\alpha].
\end{equation}
\end{proposition}
This means that  in the picture (\ref{Braiding}) representing the Barrett-Crane intertwiner, the lines are lines and not ribbons and  the
black  dot is a completely symmetric pivot.

\Proof
It is sufficient to show this property for an elementary braid, i.e of 
length one. Let $\tau$ be the transposition of the first two variables.
\begin{eqnarray}
\iota[\tau(\alpha)] P_{12}R_{12} & = 
& <\omega[\tau(\alpha)]
 \vert (\bigotimes_{i=\tau(1)}^{\tau(n)} \stackrel{\alpha_i}{\Pi})( \Delta^{(n)}(x^A)) P_{12}{\mathbb R}_{12} h(x_A)\\ \label{line1}
& = & <P_{12}\omega[\tau(\alpha)]\vert (\bigotimes_{i=1}^n \stackrel{\alpha_i}{\pi})(\tau( \Delta^{(n)})(x^A)){\mathbb R}_{12}  h(x_A) \label{line2} \\
& = &  <{\mathbb R}_{12}\omega[\alpha] \vert (\bigotimes_{i=1}^n \stackrel{\alpha_i}{\pi}) (\Delta_{}^{(n)}(x^A))  h(x_A) \label{line3} \\
& = & \iota[\alpha]\;\;.
\end{eqnarray}
We have successively used the definition of the q-BC Lorentzian intertwiner, the quasitriangular relation $\Delta_{21}={\mathbb R}_{12} \Delta_{12}
{\mathbb R}_{12}^{-1},$
 and the invariance ${\mathbb R}_{12}\omega(\alpha_1) \otimes \omega(\alpha_2)=
 \omega(\alpha_1) \otimes \omega(\alpha_2)$ shown in the appendix (\ref{Ronomega1}).
$\Box$

Remark: This property can also be shown using the decomposition in three valent
 interwiners. The fact that the action of the ${\mathbb R}$ matrix on three valent intertwiner 
amounts to multiply it by the ratio of ribbon elements and the fact that the ribbon element is trivial in a simple representation is sufficient to conclude.

\medskip

Finally, we can generalized the q-BC Lorentzian intertwiner as follows:
\begin{proposition}{The generalized q-BC Lorentzian intertwiner}

Let $\alpha=(\alpha_1,\cdots,\alpha_n)$ and $\beta=(\beta_1,\cdots,\beta_p)$ two sequences of simple representations of the quantum Lorentz group. Following (\ref{QLI}), we define the generalized q-BC intertwiner
 $\iota[\beta;\alpha] \; : \; \mathbb{V}[\alpha] \rightarrow 
 \mathbb{V}[\beta]$ by:
\begin{eqnarray} \label{GQLI}
\iota[\beta;\alpha] \; =\sum_{A,B}
 (\bigotimes_{i=1}^p \stackrel{\beta_i}{\Pi})(\Delta^{(p)}(S(x^{B})))
\vert \omega[\beta]>
 <\omega[\alpha] \vert  (\bigotimes_{j=1}^n\stackrel{\alpha_j}{\Pi})(\Delta^{(n)}(x^{A}))\; h (x_{B}x_{A})\;\;.
\end{eqnarray}

It is represented by the right graph of caption (\ref{Intertwiner}) and satisfies the following properties:
\begin{enumerate}
\item $\iota[ \beta;\alpha]$ is  an intertwiner and is well defined in the following sense:  the series   $<\stackrel{B}{e}_b(\beta)\vert \iota[\beta;\alpha] \vert \stackrel{A}{e}_a(\alpha)>$ converges. 
\item any expansion of $\iota[\beta;\alpha]$ in three valent intertwiners introduces only simple representations in intermediate channels.
\item $\iota[\beta;\alpha]$ is invariant under braiding.
$$\forall b,b'\;\;\;\iota[\beta;\sigma(b)(\alpha)]{\mathbb R}(b)=\iota[\beta;\alpha]=
{\mathbb R}(b')^{-1}\iota[\sigma(b')(\beta);\alpha].$$
\end{enumerate}
\end{proposition}
\Proof \\
This proposition is proved along the same lines as the previous propositions. $\Box$

\medskip

The graph of the right handside of caption {} as one vertex $v$ and $n$  edges entering, colored by $[\alpha]$,  and $p$ colored edges going out, colored by $[\beta]$. We will denote this graph by 
 $v[\alpha,\beta]$. It will be  useful to define  the vertex function $f_v \in Fun(AN_q) \otimes
 \text{Hom}(\mathbb{V}[\alpha],\mathbb{V}[\beta])$ by:
\begin{eqnarray}
f_v  = \sum_{A,B}
 (\bigotimes_{i=1}^p \stackrel{\beta_i}{\Pi})(\Delta^{(p)}(S(x^{B})))
\vert \omega[\beta]>
 <\omega[\alpha] \vert  (\bigotimes_{j=1}^n\stackrel{\alpha_j}{\Pi})(\Delta^{(n)}(x^{A}))\; h (x_{B}x_{A})
\end{eqnarray}
As a consequence, the Quantum Lorentzian intertwiner is recovered from the vertex function as $\iota[\beta;\alpha] \; = \; h(f_v)$. 

\section*{IV. Quantum Lorentzian Simple Spin Networks}
In this section we will define what is meant by evaluation of a closed simple spin network for the quantum Lorentz group. Rather than studying the property of convergence in general we will concentrate on  the evaluation of particular simple closed graph which are the basic ingredients for the construction of a spin foam amplitude.
In particular we will give a formula for 
 the weight associated to a 4-simplex, the so-called quantum $10j$-symbol.

Of course, as in the classical case, problems of divergences due to the non-compactness of the quantum Lorentz group appear as soon as one tries to evaluate naively a closed spin network. This difficulty is handled  by using  a regularization similar to the classical case. Moreover, quantum groups introduce problems of  non trivial braidings in the definition of spin networks which have to be handled as well.

\subsection*{IV.1. Evaluation of closed simple spin networks}
Evaluation of simple closed spin network has been studied in \cite{BC1} in the case of $SO(4)$. In the case of $SO_q(4)$ it has been studied in \cite{Ye}. Later on a study of simple closed spin network in 
the case  where the group is the Lorentz group has been done in \cite{BC2}. The
 property of integrability of simple spin network has been unraveled in 
\cite{BB2}.

Evaluation of simple spin networks, in the case of the quantum Lorentz group, contains therefore the two difficulties: 

-the network is a graph embedded in three space and its projection on a plane contains possible crossings

- the integrability, in the sense of Baez and Barett, has to be studied baring in mind that 
in  the quantum case property of convergences are usually enhanced.

 The Barrett-Crane intertwiner being formulated as an integral on the 
hyperboloid, the evaluation of a simple spin network is given by a multiple integral on a product of  hyperboloid.
Let 
\begin{eqnarray} \label{classicalkernel}
K_{\rho}(x,y) \; = \; \frac{\sin \rho l(x,y)}{\rho \sinh l(x,y)}\;\;
\end{eqnarray} 
where $x$ and $y$ are two points in the hyperboloid ${\cal H}_+$, $l(x,y)$ the distance between them and $(0,\rho)$ the simple representation coloring the edge linking $x$ to $y$. A simple spin network is an unoriented graph $\Gamma$ where the edges are colored by simple representations. We denote $\Gamma_0=\{v_1,...,v_n\}$ the set of vertices of $\Gamma$ and $\Gamma_1$ the set of edges of   $\Gamma$. 
If $\epsilon$ is an edge it defines a pair of vertices $\{v(+\epsilon),v(-\epsilon)\}.$

The evaluation of a simple spin network $\Gamma$ is given by the integral:
\begin{equation}
I(\Gamma)=
\int_{{\cal H}_+^{\times n }}
\prod_{\epsilon\in \Gamma_1}K_{\rho_\epsilon}(x_{v(+\epsilon)}, x_{v(-\epsilon)})
\prod_{v\in\Gamma_0}  dx_v.
\end{equation}

However because of the non-compactness of the Lorentz group, such an integral is
  divergent. To regularize it, one chooses any vertex $v_k$, and define $ I(\Gamma)$ as being 
the integral
$$ I(\Gamma)=
\int_{{\cal H}_+^{\times (n-1) }}
\prod_{\epsilon\in \Gamma_1}K_{\rho_\epsilon}(x_{v(+\epsilon)}, x_{v(-\epsilon)})
\prod_{v\not=v_k} dx_v.$$

It can be shown \cite{BB2} that this integral, if it converges, does not depend on $x_k,$ and does not depend on the chosen $v_k.$

\medskip

 Quantum Lorentzian spin networks are defined in an analogous way.
From an algebraic point of view our construction  is the same as Yetter's one \cite{Ye} because the quantum Lorentz algebra $U_q(so(3,1))$ is twist equivalent, as a Hopf algebra, to 
$U_q(sl(2))\otimes U_{q^{-1}}(sl(2)).$ The only, but essential, difference is contained in the star structure and in the category of representation which is considered.

A simple Lorentzian q-spin network is a  graph $\Gamma$ embedded in ${\mathbb R}^3$ colored by simple representations $[\alpha]$ of $U_q(so(3,1))$. 

The operation of evaluation of this network  is defined in the sequel and when it is finite is an invariant under isotopy.

We first define the evaluation  formally, along the lines of Yetter.
We first project the graph $\Gamma$ on a plane keeping track of the under-over crossings.
We then apply the method of Reshetikhin-Turaev \cite{RT} to compute the evaluation, where the coupons are q-BC intertwiner and the ribbons  are colored by simple representations. Note that the structure 
of ribbons is invisible because the ribbon element on simple representations is equal to $1.$

We will consider only graphs which are at least trivalent for each vertex, this implies that the q-BC intertwiner is well defined for each vertex. Unfortunately the definition of Reshetikhin-Turaev  invariant uses traces over the vector spaces  associated to the representations, therefore as such it is 
ill-defined. 

The method of regularization of this evaluation is similar to the classical one. We associate to each vertex $v$ the vertex function   $f_v,$ we tensor them
 and pair the spaces ``\`a la Reshetikhin-Turaev'' while keeping track of the braiding by introducing ${\mathbb R}$ matrices.
The obtained result is an element $f(\Gamma)\in (H_{+q}^3)^{\otimes n}$ where $n$ is the   the number of vertices. 
The formal evaluation of $\Gamma$  would be  the  integral  $h^{\otimes n}(f(\Gamma))$, 
while the regularization consists in picking an arbitrary vertex and integrating $f(\Gamma)$ over the quantum hyperboloid associated to the remaining vertices.
The invariance of the integral under $U_q(so(3,1))$ implies first that  
 $(h^{\otimes{n-1}}\otimes id)(f(\Gamma))=I(\Gamma) 1$ and secondly that this $I(\Gamma)$ is independant of the choice of the removed vertex.
When $I(\Gamma)$ is finite $\Gamma$ is said to be integrable.

We have used the following convention for evaluating simple q-spin networks. We select a projection of the graph in such a way that all the vertices are aligned and  we enumerate them from the left  to the right.

We first define the quantum analog of $K_{\rho}(x,y).$
We will define it as   an element of $(Fun_(AN_q))^{\otimes 2}$ which is 
$(id \otimes S)\Delta(\stackrel{\alpha}{K}).$
We will use the convenient abuse of notation of viewing this as a non commutative function of two variables  $\stackrel{\alpha}{K}(x,y).$ 

Using the Sweedler notation, $\Delta(a)=a_{(1)}\otimes a_{(2)},$ we obtain that
 \begin{equation}
\stackrel{\alpha}{K}(x,y)=\stackrel{\alpha}{K}_{(1)}\otimes 
S(\stackrel{\alpha}{K}_{(2)}).
\end{equation}

This definition is a direct consequence of the definition of the q-BC interwiner and of the relation
\begin{eqnarray}
\stackrel{\alpha}{K}(x,y)&=&\sum_{A,a,I,J}
<\omega(\alpha)\vert \stackrel{\alpha}{\Pi}(x^{I})\vert e^{A}_{a}>
( x_I \otimes 1) 
< e^{A}_{a}\vert  \stackrel{\alpha}{\Pi}(S(x^{J}))\vert \omega(\alpha)>
(1\otimes x_J)\nonumber\\
&=&\sum_{I,J}<\omega(\alpha)\vert \stackrel{\alpha}{\Pi}(x^I S(x^J)) \vert \omega(\alpha)>x_I\otimes x_J\nonumber\\
&=&\stackrel{\alpha}{K}_{(1)}\otimes S(\stackrel{\alpha}{K}_{(2)}).\nonumber
\end{eqnarray}

Using the definition of the coproduct (\ref{coprodfunc}), we obtain:

\begin{equation}
\stackrel{\alpha}{K}(x,y)=(id\otimes S)\sum_{I,J,K}\Lambda^{II}_{00}(\alpha)\Phi^{JK}_I\Psi_{JK}^I.
\end{equation}

 This function satisfies the following convolution property:
\begin{eqnarray}
\int \stackrel{\alpha}{K}(x,y)
\stackrel{\beta}{K}(y,z)dy = 
\; \frac{\delta(\alpha-\beta)}{P(\alpha)} \stackrel{\alpha}{K}(x,z)
\end{eqnarray}
where $P(\alpha)$ is the Plancherel measure and $\int$ is the integral over the quantum Hyperboloid. This property is a direct consequence of (\ref{SchurLemma}).

\medskip

First, let us consider the simple network  $\Gamma_2[\alpha]$ with two vertices linked by a set of $p$ edges colored by $[\alpha]=(\alpha_1,\cdots,\alpha_p)$. 
\begin{figure} 
\psfrag{a1}{$\alpha_1$}  
\psfrag{a2}{$\alpha_2$}  
\psfrag{an}{$\alpha_p$}  
\centering 
\includegraphics[scale=0.8]{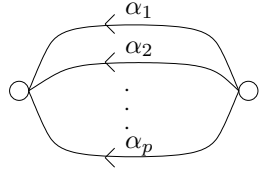} 
\caption{The graph function associated to $\Gamma_2$.}  
\label{Gamma2} 
\end{figure} 
It is straightforward to see that the associated graph function, represented by the graph (\ref{Gamma2}), is given by:
\begin{eqnarray} \label{twopointsfunction}
f({\Gamma_2}) \; = \stackrel{\alpha_1}{K}_{(1)} \cdots \stackrel{\alpha_p}{K}_{(1)}\;\;\otimes \; S(\stackrel{\alpha_1}{K}_{(2)} \cdots \stackrel{\alpha_p}{K}_{(2)}).
\end{eqnarray}
The evaluation of this spin-network, defines the so-called $\Theta_p$ functions, and are given   by:
\begin{eqnarray}
&&I(\Gamma_2[\alpha])=\Theta^q_p(\rho_1,...,\rho_p)= h(\prod_{j=1}^p \stackrel{\alpha_i}{K})
= \sum_M [d_M]_q^2 \prod_{i=1}^p \Lambda^{MM}_{\;\;0}(\alpha_i)=\\ 
&&=\sum_{n=1}^{+\infty}\frac{1}{[n]_q^{p-2}}\prod_{j=1}^p 
\frac{\sin(2n\nu_j)}{\sin(2\nu_j)}=G^q_p(\nu_1,...,\nu_p),\label{Gqp}
\end{eqnarray}
where we have denoted $\nu_i= \frac{1}{2}\kappa\rho_i.$

We will need properties of these series and will concentrate on a precise study of these series when $p=3$ and $p=4.$
These series have appeared in other contexts.

When $q=1$, $G^q_p(\nu_1,...,\nu_p)$ is, up to a constant, the value found by Witten \cite{Wi2} of  the volume of the moduli space
 ${\mathfrak{M}} (C_1,...,C_p)$ of flat $SU(2)$ connections on the sphere with $p$ punctures associated to conjugacy classes
 $C_j$, with $\exp(i\nu_j T)\in C_j$ where  $T=diag(1,-1).$

If we fix the variables  $\rho_j$ and study the behaviour $\kappa$ goes to $0$, which amounts to recover the ordinary Lorentzian Barrett-Crane model, these series  are equivalent, up to a normalization of $\kappa$, to the integrals
\begin{equation}
\int_{0}^{+\infty}\prod_{j=1}^p\frac{sin(\rho_j x)}{\rho_j} \sinh (x)^{2-p}
\end{equation} 
considered by Barrett and Crane.

When $p=3$, $G_3^q(\nu_1,\nu_2,\nu_3)$ has appeared naturally in the work \cite{BR2}, and it is precisely the definition of the coefficient  $M(\alpha_1,\alpha_2,\alpha_3).$

The functions $G^q_p(\nu_1,...,\nu_p)$ have also been  introduced by one of us  and A.Szenes in \cite{RS}, as the value of cyclic traces on non commutative deformations of  ${\mathfrak{M}} (C_1,...,C_p)$.

When $0<q<1$, 
it is trivial to show that these series are uniformally convergent for $p\geq 3$,
whereas it is a distribution in the case where $p=2$ and is equal to $\frac{\delta(\alpha_1-\alpha_2)}{P(\alpha_1)}.$

As a result, when $p\geq 3$ and $0<q<1$  the functions $G^q_p$ are continuous  functions (even smooth functions) of $\nu_1,...,\nu_p.$ 

The functions $G^q_3$ and $G^q_4$ can be expressed in term of Jacobi $\vartheta$ functions as shown in the appendix A.2. 

In particular for  $p=3,$ we obtain that the evaluation of this  simple spin network  is equal to:
\begin{eqnarray}
G_3^q(\nu_1,\nu_2,\nu_3) && =  \frac{q^{-1}-q}{16} \frac{1}{\sin(2\nu_1) \sin(2\nu_2) \sin(2\nu_3)} ( \frac{\vartheta_4'}{\vartheta_4}(\nu_1+\nu_2+\nu_3) \nonumber \\
&& -  \frac{\vartheta'_4}{\vartheta_4}(-\nu_1+\nu_2+\nu_3) -  \frac{\vartheta'_4}{\vartheta_4}(\nu_1-\nu_2+\nu_3) -  \frac{\vartheta_4'}{\vartheta_4}(\nu_1+\nu_2-\nu_3) ).
\end{eqnarray}

In particular we can recover the classical behaviour, when $\kappa\rightarrow 0$, computed in \cite{BC2}. From the modular properties of $\vartheta_4$ we 
 have for any real variable $x$:
\begin{eqnarray}
\vartheta_4(\kappa   x)  \sim  2 e^{-\frac{\pi^2}{4\kappa}} \cosh (\frac{\pi}{2}x).
\end{eqnarray}
As a result we get 
\begin{eqnarray}
\Theta_3^q(\rho_1,\rho_2,\rho_3)  & \sim & \frac{f(\kappa)}{\rho_1 \rho_2 \rho_3}  (\tanh\frac{\pi}{2}(\rho_1 + \rho_2 - \rho_3) + \tanh\frac{\pi}{2}(\rho_1 - \rho_2 + \rho_3) \nonumber \\
&& + \tanh\frac{\pi}{2}(-\rho_1 + \rho_2 + \rho_3) - \tanh\frac{\pi}{2}(\rho_1 + \rho_2 + \rho_3))
\end{eqnarray}
where $f$ is a function of $\kappa$.

The evaluation of $\Theta_4^q(\rho_1,\rho_2,\rho_3,\rho_4)$ is crucial for the computation of the weight of the spin foam on edges.
We can compute it exactly (see the appendix A.2) and we have the $q-$analogue of the formula of Barrett-Crane:
\begin{eqnarray}
G_4^q(\nu_1,\nu_2,\nu_3,\nu_4)=
\frac{(q-q^{-1})^2}{256\prod_{j=1}^4 \sin(2\nu_j)}
\sum_{\epsilon_1,\epsilon_2,\epsilon_3,\epsilon_4=\pm 1}
\epsilon_1\epsilon_2\epsilon_3\epsilon_4
\frac{\vartheta_1''}{\vartheta_1}(\sum_{i=1}^4\epsilon_i\nu_i).
\end{eqnarray}

When $q=1$,   $G^q_p(\nu_1,...,\nu_p)$, being volume of a symplectic manifold, is non negative. However it can be equal to zero, which arizes when the manifold is empty. 
Let us define the function $Y:({\mathbb R}^+)^3\rightarrow \{0,1\}$, which is the characteristic function of the  set $\Delta$ defined by
$$\Delta=\{ (a,b,c)\in ({\mathbb R}^+)^3, a\leq b+c, b\leq a+c,c\leq a+b\}.$$
And for $x\in {\mathbb R}$, denote by $\underline x\in 
[-\frac{\pi}{2}, \frac{\pi}{2}[$ the unique real number satisfying 
$x-\underline x\in \pi{\mathbb Z}.$

When $p=3$ we have 
\begin{equation}
G_3^1(\nu_1,\nu_2,\nu_3)=
\frac{\pi}{4\vert\sin(2\nu_1)\sin(2\nu_2)\sin(2\nu_3)\vert}
Y(\vert\underline{\nu_1}\vert,\vert\underline{\nu_2}\vert,
\vert\underline{\nu_3}\vert).
\end{equation}
 
The behaviour of $G_3^q$ is different, because it is always strictly positive.
Indeed, in the appendix we have shown that $G_3^q(\nu_1,\nu_2,\nu_3)\geq m$ with $m=\frac{q^{-1}-q}{16}\vartheta_2^4(0)\vartheta_4^4(0).$

We will need a similar result for  $G^q_4(\nu_1,...,\nu_4).$ 
We will use the following formula, which is trivial from the definition of
 $G^q_4(\nu_1,...,\nu_4)$ as series:
\begin{equation}
 G^q_4(\nu_1,...,\nu_4)=
\frac{1}{\pi}\int_{0}^{\pi}
 G^q_3(\nu_1,\nu_2,\nu) G^q_3(\nu,\nu_3,\nu_4)\sin^2(2\nu)d\nu.
\end{equation}
 
(Note that when q=1, this equation follows from the fact that  the moduli space of the sphere with 4 punctures is obtained by gluing the moduli spaces of two three punctured spheres.).
As a result, using the lower bound for $ G^q_3$, we obtain
$$G^q_4(\nu_1,...,\nu_4)\geq \frac{m^2}{2}.$$
This has the far reaching consequence that the function
 $\Theta_4^q(\rho_1,\rho_2,\rho_3,\rho_4)$  is bounded from below by
   $\frac{m^2}{2}.$
This behaviour is very different from the case where $\kappa=0$.
In the Riemannian case, $\rho_i$ are positive half-integers, and $\Theta_4(\rho_1,\rho_2,\rho_3,\rho_4)$ can vanish if the selection rules on $(\rho_1,\rho_2,\rho_3,\rho_4)$ are not satisfied. 
In the Lorentzian Barrett-Crane model,  $\rho_i$ are any positive number, and the infimum of 
$\Theta_4(\rho_1,\rho_2,\rho_3,\rho_4)$ is $0$, though it is never reached.

\subsection*{IV.2. Amplitude for the 4-simplex}

We will now give an expression, in term of multiple series, of  the evaluation of a colored  4-simplex (the 10j symbol).
Important properties will show up on this expression:

-it is finite

-it is a continuous function of $(\rho_1,...,\rho_{10})$

-it does not depend on the choice of crossing.

\begin{figure} 
\psfrag{a1}{$\alpha_1$}  
\psfrag{a2}{$\alpha_2$}  
\psfrag{a3}{$\alpha_3$}  
\psfrag{a4}{$\alpha_4$} 
\psfrag{a5}{$\alpha_5$}  
\psfrag{a6}{$\alpha_6$}  
\psfrag{a7}{$\alpha_7$}  
\psfrag{a8}{$\alpha_8$}
\psfrag{a9}{$\alpha_9$}  
\psfrag{a10}{$\alpha_{10}$}  
\centering 
\includegraphics[scale=0.8]{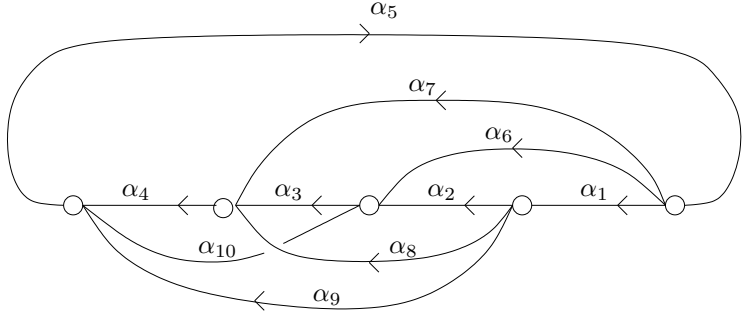} 
\caption{The $\Gamma_5^+$ spin network.}  
\label{10jsymbol} 
\end{figure}

Let us consider the graph embedded in  $\mathbb{R}^4$, whose vertices and edges are those of  the $4$ simplex  colored by a sequence $\alpha=(\alpha_1, \cdots, \alpha_{10})$ of simple representations.
When one projects this graph on a hyperplane there is a choice of crossing which has to be made. We will denote $\Gamma_5^{\pm}$ the associated simple spin network.
The following computation is done with the choice of upper crossing which is depicted in the figure (\ref{10jsymbol}).

With the convention of the picture, it is straightforward to compute the graph function:
\begin{eqnarray*}
&&f(\Gamma_5^{+})=
\sum_{A,B} \stackrel{\alpha_1}{K}(x^{A_1}S(x^{B_1}))
\stackrel{\alpha_2}{K}(x^{A_2}S(x^{B_2}))
\stackrel{\alpha_3}{K}(x^{A_3}S(x^{B_3}))
\stackrel{\alpha_4}{K}(x^{A_4}S(x^{B_4}))\\
&&\stackrel{\alpha_5}{K}(x^{A_5}\mu^{-1}S(x^{B_5}))
\stackrel{\alpha_6}{K}(x^{A_6}S(x^{B_6}))
\stackrel{\alpha_7}{K}(x^{A_7}S(x^{B_7}))\\
&&\stackrel{\alpha_8}{K}(x^{A_8}{\mathbb R}_{2}S(x^{B_8}))
\stackrel{\alpha_9}{K}(x^{A_9}S(x^{B_9}))
\stackrel{\alpha_{10}}{K}(x^{A_{10}}{\mathbb R}_{1}S(x^{B_{10}}))\\
&&x_{B_5}x_{A_4}x_{A_{10}}x_{A_9}\otimes
x_{B_4}x_{A_7}x_{A_3}x_{A_8}\otimes
x_{B_{10}}x_{B_3}x_{A_6}x_{A_{2}}\otimes
x_{B_9}x_{B_8}x_{B_2}x_{A_1}\otimes
x_{B_1}x_{B_{6}}x_{B_7}x_{A_9},
\end{eqnarray*}
where we have abusively denoted ${\mathbb R}_{12}={\mathbb R}_{1}\otimes
{\mathbb R}_{2}.$

Using the properties $\mu \omega(\alpha)=\omega(\alpha)$ and 
$S(\stackrel{\alpha}{K})\stackrel{\alpha}{K},$
 this function can be written as:
\begin{eqnarray*}
&&f(\Gamma_5^{+})=
\stackrel{\alpha_{5}}{K}_{(1)}
\stackrel{\alpha_{4}}{K}_{(1)}
\stackrel{\alpha_{10}}{K}_{(1)}
\stackrel{\alpha_{9}}{K}_{(1)}
\otimes 
S(\stackrel{\alpha_{4}}{K}_{(2)})
\stackrel{\alpha_{7}}{K}_{(1)}
\stackrel{\alpha_{3}}{K}_{(1)}
\stackrel{\alpha_{8}}{K}_{(1)}
\otimes
S(\stackrel{\alpha_{10}}{K}_{(3)})
S(\stackrel{\alpha_{3}}{K}_{(2)})
\stackrel{\alpha_{6}}{K}_{(1)}
\stackrel{\alpha_{2}}{K}_{(1)}
\otimes\\
&&S(\stackrel{\alpha_{9}}{K}_{(2)})
S(\stackrel{\alpha_{8}}{K}_{(3)})
S(\stackrel{\alpha_{2}}{K}_{(2)})
\stackrel{\alpha_{1}}{K}_{(1)}
\otimes
S(\stackrel{\alpha_{1}}{K}_{(2)})
S(\stackrel{\alpha_{6}}{K}_{(2)})
S(\stackrel{\alpha_{7}}{K}_{(2)})
S(\stackrel{\alpha_{5}}{K}_{(2)})
<\stackrel{\alpha_{8}}{K}_{(2)}, {\mathbb R}_{2}><\stackrel{\alpha_{10}}{K}_{(2)}, {\mathbb R}_{1}>.
\end{eqnarray*}
We evaluate this spin network by applying $(id\otimes h^{\otimes 4}).$
We obtain:
\begin{eqnarray*}
&&(id\otimes h^{\otimes 4})(f(\Gamma_5^+))=
\stackrel{\alpha_{5}}{K}_{(1)}
\stackrel{\alpha_{4}}{K}_{(1)}
\stackrel{\alpha_{10}}{K}_{(1)}
\stackrel{\alpha_{9}}{K}_{(1)} 
h(S(\stackrel{\alpha_{4}}{K}_{(2)})
\stackrel{\alpha_{7}}{K}_{(1)}
\stackrel{\alpha_{3}}{K}_{(1)}
\stackrel{\alpha_{8}}{K}_{(1)})
h(
S(\stackrel{\alpha_{10}}{K}_{(3)})
S(\stackrel{\alpha_{3}}{K}_{(2)})
\stackrel{\alpha_{6}}{K}_{(1)}
\stackrel{\alpha_{2}}{K}_{(1)})
\\
&&h(S(\stackrel{\alpha_{9}}{K}_{(2)})
S(\stackrel{\alpha_{8}}{K}_{(3)})
S(\stackrel{\alpha_{2}}{K}_{(2)})
\stackrel{\alpha_{1}}{K}_{(1)})
h(
\stackrel{\alpha_{5}}{K}_{(2)}
\stackrel{\alpha_{7}}{K}_{(2)}
\stackrel{\alpha_{6}}{K}_{(2)}
\stackrel{\alpha_{1}}{K}_{(2)})
<\stackrel{\alpha_{8}}{K}_{(2)}, {\mathbb R}_{2}><\stackrel{\alpha_{10}}{K}_{(2)}, {\mathbb R}_{1}>.
\end{eqnarray*}

Using the invariance on the left and on the right of $h$ and the braiding relation,
we obtain $(id\otimes h^{\otimes 4})(f(\Gamma_{5}^{+}))=
I(\Gamma_{5}^{+}) 1$ where 
\begin{eqnarray*}
&&I(\Gamma_{5}^{+})=
h(S(\stackrel{\alpha_{4}}{K}_{})
\stackrel{\alpha_{7}}{K}_{(1)}
\stackrel{\alpha_{3}}{K}_{(1)}
\stackrel{\alpha_{8}}{K}_{(1)})
h(
S(\stackrel{\alpha_{10}}{K}_{(2)})
S(\stackrel{\alpha_{3}}{K}_{(2)})
\stackrel{\alpha_{6}}{K}_{(1)}
\stackrel{\alpha_{2}}{K}_{(1)})
\\
&&h(S(\stackrel{\alpha_{9}}{K}_{})
S(\stackrel{\alpha_{8}}{K}_{(3)})
S(\stackrel{\alpha_{2}}{K}_{(2)})
\stackrel{\alpha_{1}}{K}_{(1)})
h(
\stackrel{\alpha_{5}}{K}_{}
\stackrel{\alpha_{7}}{K}_{(2)}
\stackrel{\alpha_{6}}{K}_{(2)}
\stackrel{\alpha_{1}}{K}_{(2)})
<\stackrel{\alpha_{8}}{K}_{(2)}, {\mathbb R}_{2}><\stackrel{\alpha_{10}}{K}_{(1)}, {\mathbb R}_{1}>.
\end{eqnarray*}

We eliminate the ${\mathbb R}$ matrix using the property (\ref{Ronomega2}), and the expression reduces to:
\begin{eqnarray}
&&I(\Gamma_{5}^{+})=
h(S(\stackrel{\alpha_{4}}{K}_{})
\stackrel{\alpha_{7}}{K}_{(1)}
\stackrel{\alpha_{3}}{K}_{(1)}
\stackrel{\alpha_{8}}{K}_{(1)})
h(
S(\stackrel{\alpha_{10}}{K}_{})
S(\stackrel{\alpha_{3}}{K}_{(2)})
\stackrel{\alpha_{6}}{K}_{(1)}
\stackrel{\alpha_{2}}{K}_{(1)})\nonumber
\\
&&h(S(\stackrel{\alpha_{9}}{K}_{})
S(\stackrel{\alpha_{8}}{K}_{(2)})
S(\stackrel{\alpha_{2}}{K}_{(2)})
\stackrel{\alpha_{1}}{K}_{(1)})
h(
\stackrel{\alpha_{5}}{K}_{}
\stackrel{\alpha_{7}}{K}_{(2)}
\stackrel{\alpha_{6}}{K}_{(2)}
\stackrel{\alpha_{1}}{K}_{(2)}).\label{evaluationof4simplex}
\end{eqnarray}

It remains to evaluate the four integrals.
We denote by $\stackrel{A}{E}{}^i_j$ the basis of 
$Mat_{2A+1}(\mathbb{C})$  which satisfies
 $\stackrel{A}{E}{}^i_j
 \stackrel{A}{E}{}^k_l=\delta_j^k 
\stackrel{A}{E}{}^i_l.$

As a result, in this basis, (\ref{coprodfunc}) is expressed as:
\begin{equation}
\Delta(\stackrel{\alpha}{K})=
\sum_{I,J,K}\Lambda^{II}_{00}(\alpha)
\Clebphi{J}{K}{I}{j}{k}{i}
\Clebpsi{J}{K}{I}{j'}{k'}{i}
\stackrel{J}{E}{}^{j'}_j\otimes 
\stackrel{K}{E}{}^{k'}_k,
\end{equation}

and the action of the antipode $S$ is given by:

\begin{equation}
S(\stackrel{A}{E}{}^{i}_j) =
\stackrel{A}{w}_{jn}\stackrel{A}{E}{}^n_m(\stackrel{A}{w}{}^{-1}){}^{mi}.
\end{equation} 

where $\stackrel{A}{w}$ is the matrix  defined as :

\begin{equation}
\stackrel{A}{w}_{mn}=[d_A]^{1/2}e^{-i\pi A}\Clebpsi{A}{A}{0}{m}{n}{0}.
\end{equation} 

\begin{figure} 
\psfrag{M1}{$M_1$}  
\psfrag{M2}{$M_2$} 
\psfrag{M3}{$\!\!\!M_3$}  
\psfrag{M4}{$M_4$}
\psfrag{M5}{$M_5$}  
\psfrag{M6}{$M_6$} 
\psfrag{M7}{$M_7$}  
\psfrag{M8}{$M_8$}
\psfrag{M9}{$M_9$}  
\psfrag{M10}{$M_{10}$} 
\centering 
\includegraphics[scale=0.6]{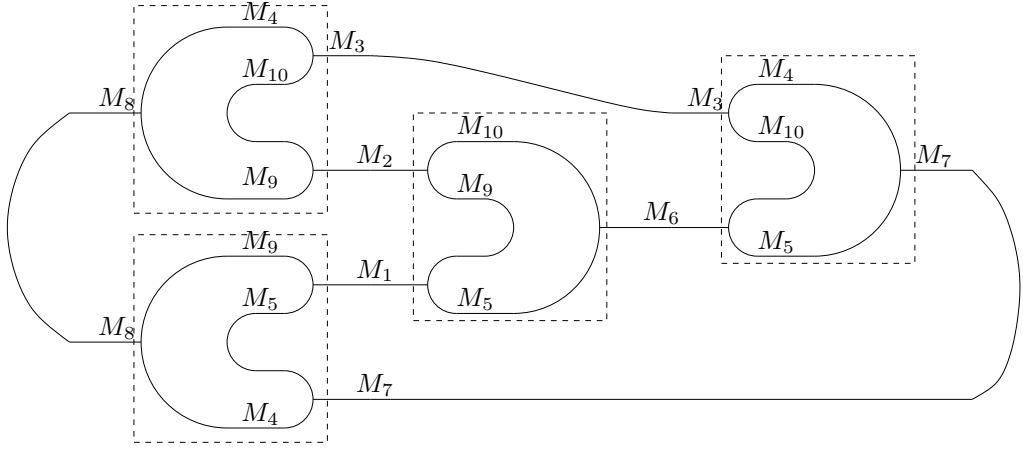} 
\caption{The spin network associated to the function $X(M_1,...,M_{10})$.}  
\label{Graph} 
\end{figure}

 It is then straighforward  to perform the integrations of
 (\ref{evaluationof4simplex}), and we get:

\begin{equation}
I(\Gamma_5^{+})=\sum_{M_1,\cdots,M_{10}}
\prod_{i=1}^{10} \Lambda^{M_i M_i}_{\;\;00} (\alpha_i)
[d_{M_4}]_q[d_{M_5}]_q[d_{M_9}]_q[d_{M_{10}}]_q X(\{M_i\})
\end{equation}
where the function $X$ is the  evaluation of the $U_q(su(2))$ spin 
network drawn in picture (\ref{Graph}).

We now evaluate this spin network.
Each box is replaced by the corresponding $3j$ intertwiner multiplied by a number. For example the box on the upper left is replaced by
 $\Psi^{M_3M_2}_{M_8}\lambda(M_3,M_2,M_8;M_4,M_{10},M_9)$ where 
\begin{equation}
\lambda(M_3,M_2,M_8;M_4,M_{10},M_9)=\frac{[d_{M_3}]_q^{1/2}}{[d_{M_4}]_q^{1/2}}
\sixj{M_9}{M_8}{M_4}{M_3}{M_{10}}{M_2}
.
\end{equation}
As a result we obtain:
\begin{eqnarray}
&&X=\frac{[d_{M_1}]_q[d_{M_3}]_q^{1/2}[d_{M_6}]_q^{1/2}[d_{M_7}]_q^{1/2}[d_{M_8}]_q^{1/2}}
{[d_{M_4}]_q^{1/2}[d_{M_5}]_q[d_{M_9}]_q^{1/2}}\sixj{M_3}{M_7}{M_6}{M_1}{M_2}{M_8} \sixj{M_4}{M_7}{M_5}{M_6}{M_{10}}{M_3}\nonumber\\
&&\sixj{M_{10}}{M_6}{M_5}{M_1}{M_9}{M_{2}} \sixj{M_4}{M_8}{M_9}{M_1}{M_5}{M_7}
 \sixj{M_9}{M_8}{M_4}{M_3}{M_{10}}{M_2}.
\end{eqnarray}

Using Racah-Wigner symmetry, we finally obtain   the nice formula:
\begin{eqnarray}
I(\Gamma_5^{+}) & = & \sum_{M_1,\cdots,M_{10}} \left( \prod_{i=1}^{10} [d_{M_i}]_q^{1/2} \Lambda^{M_i M_i}_{\;\;00} (\alpha_i) \right) \sixj{M_3}{M_7}{M_6}{M_1}{M_2}{M_8} \sixj{M_4}{M_7}{M_5}{M_6}{M_{10}}{M_3} \nonumber \\
&& \sixj{M_2}{M_6}{M_1}{M_5}{M_9}{M_{10}} \sixj{M_4}{M_8}{M_9}{M_1}{M_5}{M_7} \sixj{M_9}{M_8}{M_4}{M_3}{M_{10}}{M_2}\label{4simplexevaluated}\;\;. 
\end{eqnarray}

This expression is a multiple series which converges absolutely.
 Indeed,  each $q6j$-symbol is bounded by one and when $M$ goes to infinity,  $[d_{M}]_q^{1/2} \Lambda^{M M}_{\;\;00} (\alpha) = {\cal O}(Mq^M)$. As a result the graph $\Gamma_5^{+}$ is integrable.
Because the convergence is uniform in $\rho_j$, $I(\Gamma_5^{+})$ is a continuous function of the variables $\rho_j$.

Note a very important property satisfied by $I(\Gamma_5^{+})$: it is a function of $q$ symmetric in the exchange $q \mapsto q^{-1}.$ This is a direct consequence of the invariance of the $6j$ symbols under the exchange $q\mapsto q^{-1}.$

The evaluation of the graph $\Gamma_5^{-}$ proceeds along the same lines, and its evaluation gives the same answer (\ref{4simplexevaluated}). This is compatible with the invariance $q\mapsto q^{-1}.$

\section*{V. q-Lorentzian Spin Foam Models}
With all the tools introduced in the previous section, we can now built cosmological Lorentzian spin foam models. 
Let ${\cal T}$ be a triangulation of the Lorentzian manifold $M$, we define by $\Delta_n$ the set of $n-$simplices of the triangulation.
The partition function of the spin foam model is defined as usual:
\begin{equation}
Z({\cal T})=\int_{\rho_f\in [0, \frac{2\pi}{\kappa}]}\prod_{f\in \Delta_2} d\rho_f \prod_{f\in \Delta_2}
A_2 (\rho_f)
 \prod_{t\in \Delta_3 }A_3(\rho_1^t,...,\rho_4^t ) 
\prod_{s\in \Delta_4}A_4(\rho_1^s,...,\rho_{10}^s ),
\end{equation}
where $A_4(\rho_1^s,...,\rho_{10}^s )$ is the value of the $10j$ symbol of the $4-$ simplex $s$. Note that the value of $A_2 (\rho_f)$ and  
$A_3(\rho_1^t,...,\rho_4^t )$ is still a matter of debate. See \cite{BCHT} for a clear survey of the situation. In this article three choices of $A_2$ and $A_3$ have been analyzed. We give their straighforward q-deformation in the sequel:

1)$A_2 (\rho_f)=P(\rho_f), 
A_3(\rho_1,...,\rho_4)=\Theta_4^q(\rho_1,...,\rho_4)^{-1}$

2)$A_2 (\rho_f)=P(\rho_f), 
A_3(\rho_1,...,\rho_4)=\Theta_4^q(\rho_1,...,\rho_4)$
 
3)$A_2(\rho_f)=1, A_3(\rho_1,...,\rho_4)=\Theta_4^q(\rho_1,...,\rho_4)^{-1},$

where $P$ is the Plancherel measure. 

It is therefore straighforward to show that the integral $Z({\cal T})$ is finite for the three previous cases. The integration is on a compact set, the function $A_4$ is continuous, and because of the lower bound on $\Theta_4^q$, the function $A_3$ is also continuous for the three cases. As a result $Z({\cal T})$ is a convergent integral for these three cases and more generally in  the case where $A_2, A_3$ are continuous functions of the $\rho$'s.

\section*{VI. Conclusion}
In our work we have defined cosmological deformation of Lorentzian spin foam models. The only analysis that we have made of this model is its finiteness property for a fixed triangulation.

Presently there are lots of work on the study of Lorentzian spin foam models and its relations to quantum gravity. Central questions are  the ``derivation'' of these models from the Lagrangian field theory, the restriction to the gravitational sector, the analysis of the asymptotic properties of the weight of a simplex, the relation to the Hamiltonian constraint, the possible non-perturbative definitions of the sum over the 2-complexes, the importance of the degenerate metrics, the nature of the spectrum of observables (continuous or discrete?).  They are all related to the following three questions: how fondamental is the Barret-Crane model? What are the observables and their spectrum? How do we do a semi-classical study of spin-foam models?

The same questions can be formulated in the context of cosmological deformation of spin foam models as well. Even more pressing  questions are present: how do we precisely connect the Plebanski Lagrangian field theory with a cosmological constant to cosmological deformation of the Barrett-Crane model? How do we compute $A_2$ and $A_3$?  How do we do a semi-classical expansion  around De Sitter space? What precisely are the states of quantum Lorentzian gravity with a positive cosmological constant? How do we couple these models to matter and extract  information about physical quantities?

The fate   of the spinfoam model approach depends crucially on the possibility of answering  to these questions.

There are of course extensions of our work which can be readily done.

The first one is the inclusion of faces of timelike type. The technology of quantum groups that we have developped in the past should give an  easy answer to this extension. 

The spinfoam model that we have defined should also appear as the perturbative expansion of a field theory on a non commutative analogue of $({\cal H}_+^3)^{\times 5}$.   

Another well studied question nowadays, in the classical case,  is the asymptotic properties of the $10j$ symbol. A similar question can be asked in the cosmological deformation case. 
Its formulation is as follows: we fix  $\tilde{\rho_j}$, define $\rho_j$ by
 $\tilde{\rho_j} =l_P^2 \rho_j$, and 
we would like to understand the behaviour of
 $A_4(\rho_1,...,\rho_{10})$ in the limit where $l_P$ goes to 0.
It is straightforward to see that 
\begin{equation}
A_4(\rho_1,...,\rho_{10})=
\sum_{M_1,\cdots,M_{10}}
  \prod_{j=1}^{10} [d_{M_j}]_q^{1/2}\frac{\sin((2M_j+1)\omega_j)}
{\sin(\omega_j)}R_q(M_1,...,M_{10})
\end{equation}

where 
$R_q(M_1,...,M_{10})$ is the product of the five $(6j_q)$ symbols and
 $\omega_j=\tilde{\rho_i}/l_C^2.$

The asymptotic properties of the simplex is therefore  the classical expansion of this series around $q=1$. The form of this series is very reminiscent of the series that we have studied in \cite{RS}, for which we have  developped tools (Residue formulas) for computing the asymptotic of such series. We hope to apply analog of these methods to analyze the behaviour of the $4$-simplex.

\subsection*{Acknowledgments}
We thank L.Crane and C.Rovelli for stimulating and enthousiastic discussions on spin foam models. We also thank E.Buffenoir for interesting discussions.

\section*{Appendix}

	\subsection*{A.1. The quantum Lorentz group}
 The Lie algebra $so(3,1)$ is also the real Lie algebra of 
$sl(2,\mathbb C).$ Therefore we will use indifferently the notation  $U_q(so(3,1))$ or $U_q(sl(2,\mathbb C))$ for its quantization.
We define the quantum Lorentz group $U_q(so(3,1))$  as being the quantum double of  $U_q(su(2))$. This construction is a standard and very important  construction in the theory of quantum groups. It is exposed in detail in \cite{BR1}. 
The quantum double of $U_q(su(2))$ is
 the Hopf algebra ${\cal D} = U_q(su(2)) \hat{\otimes} F_q(SU(2))^{op}$. 
$F_q(SU(2))^{op}$ denotes the Hopf algebra $F_q(SU(2))$ with permuted 
 coproduct.  The notation ${\cal D} = A \hat{\otimes} A^{*op}$ reminds the reader that inside ${\cal D}$, $(A \otimes 1)$ and $(1 \otimes A^{*op})$ do not commute but satisfy braided identities.

In \cite{BR1} we have classified and given simple expressions for the unitary irreducible representations of  $U_q(so(3,1)).$
We will only need the representation of the principal series which we denote $\stackrel{\alpha}{\Pi}$ where $\alpha=(k,\rho).$
Matrix elements of $\stackrel{\alpha}{\Pi}$ are linear forms on 
 $U_q(sl(2,{\mathbb C}))$ and therefore are elements of  
$Fun(SL_q(2,{\mathbb C})).$

Using the decomposition $Fun(SL_q(2,{\mathbb C}))=Fun(SU_q(2))\otimes Fun_q(AN)$ we have expressed in \cite{BR1} the matrix elements of  $\stackrel{\alpha}{\Pi}$ in terms of a basis of  $Fun(SU_q(2))\otimes Fun_q(AN).$
We will use the following convenient basis which elements are 
$\stackrel{A_1}{k}{}^{a_1}_{a_1'}\otimes \stackrel{A_2}{E}{}^{a_2}_{a_2'}$. We will denote this basis by $(x_A)$.
The dual $(Fun(SL_q(2,{\mathbb C}))^{*}$ is  $U_q(sl(2,{\mathbb C}))$ (more precisely contains $U_q(sl(2,{\mathbb C})$ as  a Hopf sub-algebra). 
One of its basis is the dual basis  $(x^A)$ which is the basis 
$\stackrel{A_1}{F}{}^{a_1}_{a_1'}\otimes \stackrel{A_2}{g}{}^{a_2}_{a_2'}$
with the duality bracket given by:
\begin{equation}
<\stackrel{A_1}{F}{}^{a_1}_{a_1'}\otimes \stackrel{A_2}{g}{}^{a_2}_{a_2'}
\vert
\stackrel{B_1}{k}{}^{b'_1}_{b_1}\otimes \stackrel{B_2}{E}{}^{b_2'}_{b_2}>=
\delta_{A_1B_1}\delta_{A_2B_2}\delta_{a_1b_1}
\delta_{a_1'b_1'}\delta_{a_2b_2}\delta_{a_2'b_2'}
\end{equation}

We have given explicit formulas for the representation 
$\stackrel{\alpha}{\Pi}$ in this basis. Namely:
\begin{eqnarray}
\stackrel{\alpha}{\Pi}(\stackrel{A}{F}{}^{a}_{a'}\otimes 
\stackrel{B}{g}{}^{b}_{b'})\stackrel{C}{e}_{c}=
\stackrel{A}{e}_{a'}\sum_{D}\Lambda^{BD}_{AC}(\alpha)
\Clebphi{A}{B}{D}{a}{b}{d}
\Clebpsi{B}{C}{D}{b'}{c}{d}
\end{eqnarray}
 where 
$\Lambda^{BD}_{AC}(\alpha)$ are coefficients defined in terms of analytic continuation of $6j$ symbols of $U_q(su(2))$ and whose properties are studied in depth in \cite{BR1}. An explicit expression is given in equation (89) of \cite{BR1}.
As a result we get the expression:
\begin{equation}
<\stackrel{A}{e}_a\vert \stackrel{\alpha}{\Pi}(.)\vert\stackrel{B}{e}_b>=
\sum_{C,D}\stackrel{A}{k}{}^a_{a'} \otimes \stackrel{C}{E}{}^c_{c'} 
 \Lambda^{CD}_{AB}(\alpha)\Clebphi{A}{C}{D}{a'}{c'}{d}
\Clebpsi{C}{B}{D}{c}{b}{d}.
\end {equation}

The explicit expression of $\Lambda^{CD}_{AB}(\alpha)$ simplifies drastically when  $A=B=0$ and $C=D$, and we get
\begin{equation}
\Lambda^{CC}_{00}(\alpha)=\frac{[(2C+1)i\rho]_q}{[(2C+1)]_q[i\rho]_q}.
\end{equation}

This  particular nice result is fortunate because it is this expression which is needed in the case of simple representations.

The ${\mathbb R}$ matrix of $U_q(sl(2,\mathbb{C})$ is given by the construction of the quantum double, its evaluation on principal representations has been computed in \cite{BR2}:
\begin{equation}
< \stackrel{C}{e}_c\otimes \stackrel{D}{e}_d\vert
(\stackrel{\alpha}{\Pi}\otimes\stackrel{\beta}{\Pi})({\mathbb R})
\vert \stackrel{A}{e}_a\otimes\stackrel{B}{e}_b>=
\delta_{A}^{C}\delta_a^c \sum_{M}\Lambda^{AM}_{DB}(\beta)
\Clebphi{D}{C}{M}{d}{c}{m}
\Clebpsi{A}{B}{M}{a}{b}{m}.
\end{equation}
In particular this expression implies that:
\begin{eqnarray}
(\stackrel{\alpha}{\Pi}\otimes\stackrel{\beta}{\Pi})({\mathbb R})
(\omega(\alpha)\otimes id)&=&\omega(\alpha)\otimes id\label{Ronomega1}\\
(<\omega(\alpha)\vert \otimes id)
(\stackrel{\alpha}{\Pi}\otimes\stackrel{\beta}{\Pi})({\mathbb R})&=&
<\omega(\alpha)\vert \otimes id)\label{Ronomega2}
\end{eqnarray}

which is a consequence of $\Lambda^{0B}_{BB}(\beta)=1$ when $\beta$ is simple.

In \cite{BR2} we have made a study of intertwining operators for the tensor product of principal representations.
We have shown that if $\alpha_1,\alpha_2,\alpha_3$ label principal representations, the space of intertwiners 
$\stackrel{\alpha_1}{\mathbb V}\otimes
 \stackrel{\alpha_2}{\mathbb V}\rightarrow  \stackrel{\alpha_3}{\mathbb V}$ is one dimensional when $k_1+k_2+k_3\in 2{\mathbb Z}$ and  zero 
otherwise.
We have chosen a particular normalized intertwiner, denoted
 $\Psi_{\alpha_1\alpha_2}^{\alpha_3}:
\stackrel{\alpha_1}{\mathbb V}\otimes
 \stackrel{\alpha_2}{\mathbb V}\rightarrow  \stackrel{\alpha_3}{\mathbb V}$, 
which satisfies the following completeness relation:
\begin{equation}
\|v\otimes w\|^2=\int d\beta
  M(\alpha_1,\alpha_2,\beta)  \| \Psi_{\alpha_1\alpha_2}^{\beta}(v\otimes w)\|^2,
\end{equation}
with $M(\alpha_1,\alpha_2,\beta)$ a normalization factor given by an infinite product expression and the integral $\int d\beta$ means, as usual,  an integral over
 $\rho_\beta$ and a sum over the integer $k_\beta$.

We will define $ \Phi^{\alpha_1\alpha_2}_{\alpha_3}=
(\Psi_{\alpha_1\alpha_2}^{\alpha_3})^{\dagger}.$
The complete reducibility of the tensor product 
$\stackrel{\alpha_1}{\Pi}\otimes \stackrel{\alpha_2}{\Pi}$ is expressed as:
\begin{equation}
\stackrel{\alpha_1}{\Pi} \otimes \stackrel{\alpha_2}{\Pi} = 
\int d\alpha_3  \; \Phi^{\alpha_1 \alpha_2}_{\;\;\alpha_3} \stackrel{\alpha_3}{\Pi} \; \Psi_{\alpha_1 \alpha_2}^{\;\;\alpha_3} \label{orthogonalityrelation}. 
\end{equation}

Note that when $\alpha_1,\alpha_2,\alpha_3$ are simple the expression for $M$ reduces to a simple ratio of theta functions, as explained in \cite{BR2} and recalled in the appendix A.2 of the present work. As explained in the sequel, this function $M$ is exactly the evaluation of the 
so-called ``Theta'' graph $\Theta_3.$ 

It can be shown, using Plancherel theorem \cite{BR1}, that the following identity holds as distributions:

\begin{equation}
(\stackrel{\alpha}{\Pi}\otimes\stackrel{\beta}{\Pi})(1)=
\frac{1}{P(\alpha)}  \Phi^{\alpha \beta}_{0}\Psi_{\alpha \beta}^{0}.
\label{SchurLemma}
\end{equation}
where $P(\alpha)$ is the Plancherel density, given by $P(\alpha)=
(q-q^{-1})^2\vert [i\rho] \vert^2.$

	\subsection*{A.2. Evaluation of the functions $G^q_p$}

The functions $G^q_p$ are defined by (\ref{Gqp}), they can be expressed for $p=3$ and $p=4$, in terms of Jacobi theta functions.

 Let $\tau \in \mathbb C$ and $q=\exp(i\pi \tau)$ such that $\vert q \vert <1$. We  define the complex function:
\begin{eqnarray} \label{theta}
&&\vartheta_4(z) \; = \; \sum_{n=-\infty}^{+\infty} (-1)^n q^{n^2} \exp (2inz)\;\;=\\
&&=\prod_{n=1}^{+\infty}(1-q^{2n})(1-q^{2n-1}e^{2iz})(1-q^{2n-1}e^{-2iz}).
\end{eqnarray}
The three other  Jacobi theta functions are defined  as follows:
$$
 \vartheta_1(z)  =  -i e^{i(z+\frac{\pi}{4}\tau)} \vartheta_4(z+\frac{\pi}{2} \tau) \; , \;\; \vartheta_2(z) = \vartheta_1(z+\frac{\pi}{2})\;\;, 
\vartheta_3(z) =  \vartheta_4(z+\frac{\pi}{2}).$$

These functions satisfy the fundamental Jacobi identities
\begin{eqnarray} \label{jacobi}
&&2 \vartheta_4(x') \vartheta_4(y') \vartheta_4(z') \vartheta_4(w')  =  \vartheta_1(x) \vartheta_1(y) \vartheta_1(z) \vartheta_1(w) - \vartheta_2(x) \vartheta_2(y) \vartheta_2(z) \vartheta_2(w) \nonumber \\
&+&\vartheta_3(x) \vartheta_3(y) \vartheta_3(z) \vartheta_3(w) + \vartheta_4(x) \vartheta_4(y) \vartheta_4(z) \vartheta_4(w)\;\;,
\end{eqnarray}
where $2x'=w-x+y+z$, $2y'=w+x-y+z$, $2z'=w+x+y-z$ and  $2w'=-w+x+y+z$. 
We will now choose $q=\exp(-\kappa)$, with $\kappa\in \mathbb{R^+}.$

We will first evaluate $G_3^q(\nu_1,\nu_2,\nu_3).$
Using the expansion of a product of sinuses in sum of sinuses, we obtain:
\begin{eqnarray}
&&4\prod_{j=1}^3 \sin(2\nu_j)G_3^q(\nu_1,\nu_2,\nu_3)=\\
&&F_3(-\nu_1+\nu_2+\nu_3)+F_3(\nu_1-\nu_2+\nu_3)+
F_3(\nu_1+\nu_2-\nu_3)-F_3(\nu_1+\nu_2+\nu_3)\nonumber
\end{eqnarray}

with 
\begin{equation}
F_3(z)=\sum_{n=1}^{\infty} \frac{\sin (2nz)}{[n]_q}=(q^{-1}-q)
\sum_{n=1}^{\infty} \frac{q^n \sin(2nz)}{1-q^{2n}}.
\end{equation}

To express $F_3(z)$ in term of theta functions, one can use the following trick which will be useful later on.
We first expand $\frac{1}{1-q^{2n}}=\sum_{k=0}^{+\infty}q^{2nk}$, and we 
interchange the sum in $F_3(z)$ by summing first over $k,$
we therefore obtain that 
\begin{equation}
2F_3(z)=\frac{1}{i}
\sum_{k=0}^{+\infty}
(\frac{e^{iz}q^{2k+1}}{1-e^{iz}q^{2k+1}}-
\frac{e^{-iz}q^{2k+1}}{1-e^{-iz}q^{2k+1}})=
\frac{\vartheta_4'(z)}{\vartheta_4(z)}.
\end{equation}

\medskip

As a result  we obtain:
\begin{eqnarray}
G_3^q(\nu_1,\nu_2,\nu_3) && =  \frac{q^{-1}-q}{16} \frac{1}{\sin(2\nu_1) \sin(2\nu_2) \sin(2\nu_3)} ( \frac{\vartheta_4'}{\vartheta_4}(\nu_1+\nu_2+\nu_3) \nonumber \\
&& -  \frac{\vartheta'_4}{\vartheta_4}(-\nu_1+\nu_2+\nu_3) -  \frac{\vartheta'_4}{\vartheta_4}(\nu_1-\nu_2+\nu_3) -  \frac{\vartheta_4'}{\vartheta_4}(\nu_1+\nu_2-\nu_3) ).
\end{eqnarray}

Interestingly this  series can be written as an infinite product using the fundamental Jacobi identity.

If we use the formula (\ref{jacobi}), where $x,y,z$ are replaced respectively by $2\nu_1,2\nu_2,2\nu_3$ and $w=0$, we get:
\begin{eqnarray} \label{thetaq}
G_3^q(\nu_1,\nu_2,\nu_3) & = & \frac{1}{8} \frac{\vartheta_1(2\nu_1)\vartheta_1(2\nu_2)\vartheta_1(2\nu_3)}{\sin(2\nu_1) \sin(2\nu_2) \sin(2\nu_3)} \times \\
&&\times \frac{q^{1/4}(q^{-1}-q)(q^2;q^2)^3_{\infty}}{\vartheta_4(\nu_1+\nu_2+\nu_3)\vartheta_4(-\nu_1+\nu_2+\nu_3)\vartheta_4(\nu_1-\nu_2+\nu_3)\vartheta_4(\nu_1+\nu_2-\nu_3)}\;\; \nonumber
\end{eqnarray}
where we have used the fact that $\vartheta'_1(0)= 2 q^{1/4}(q^2;q^2)^3_{\infty}$ with
\begin{eqnarray}
(q^2;q^2)_{\infty} \; = \; \prod_{k=1}^{\infty} (1-q^{2k})\;\;.
\end{eqnarray}

This enables us to show that $G_3^q(\nu_1,\nu_2,\nu_3)$ is strictly positive and that we have the lower bound:
\begin{equation}
\forall \nu_1,\nu_2,\nu_3\in {\mathbb R},G_3^q(\nu_1,\nu_2,\nu_3)\geq m\;\;\;
\text{with}\;\;m=\frac{q^{-1}-q}{16}\vartheta_2^4(0)\vartheta_4^4(0).
\end{equation}
To prove this lower bound we  use the infinite product formula for $\vartheta_4$ and $\vartheta_1$.
As a result we obtain the inequalities, when $x$ is real,:
\begin{eqnarray}
&&0\leq \vartheta_4(x)\leq \prod_{n=1}^{+\infty}(1-q^{2n})(1+q^{2n-1})^2\\
&&\frac{\vartheta_1(x)}{\sin(x)}\geq 2q^{\frac{1}{4}}
\prod_{n=1}^{+\infty}(1-q^{2n})^3.
\end{eqnarray}
As a result we obtain that $G_3^q(\alpha_1,\alpha_2,\alpha_3)\geq m$ where
 $$m=(1-q^2)\prod_{n=1}^{+\infty}\frac{(1-q^{2n})^8}{(1+q^{2n-1})^8}=
\frac{(1-q^2)}{16q}\frac{\vartheta_1'(0)^4}{\vartheta_3^4(0)}=
\frac{q^{-1}-q}{16}\vartheta_2^4(0)\vartheta_4^4(0).$$

It is easy to compute the behaviour of $m$ when $\kappa$ goes to zero.
This is done with the use of modular transformation on the theta functions and we get:
\begin{equation}
m(\kappa)\sim_{0+} \frac{2\pi^4}{\kappa^3}e^{-\frac{\pi^2}{\kappa}}.
\end{equation}

We can also express $G_4^q$ in terms of theta functions. Although we will not need these formulas, we give them here for completeness.

We have 
\begin{equation}
16\prod_{j=1}^4 \sin(2\nu_j)G_4^q(\nu_1,\nu_2,\nu_3,\nu_4)=
\sum_{\epsilon_1,\epsilon_2,\epsilon_3,\epsilon_4=\pm 1}
\epsilon_1\epsilon_2\epsilon_3\epsilon_4
F_4(\sum_{i=1}^4\epsilon_i\nu_i),
\end{equation}
where $$F_4(z)=\sum_{n=1}^{+\infty}\frac{\cos(nz)}{[n]_q^2}.$$

This last series can be evaluated as follows.
\begin{eqnarray}
&&(q-q^{-1})^2 F_4(z)=\sum_{n=1}^{+\infty}\frac{q^{2n}\cos(nz)}{(1-q^{2n})^2}\\
&&=\sum_{n=1}^{+\infty}\sum_{k=0}^{+\infty}q^{2n}\cos(nz)q^{2nk}(k+1)\\
&&=\frac{1}{2}(\sum_{k=1}^{+\infty}
\frac{ke^{iz}q^{2k}}{1-q^{2k}e^{iz}}+
\frac{ke^{-iz}q^{2k}}{1-q^{2k}e^{-iz}}).
\end{eqnarray}
From the expression of $\vartheta_1$ as an infinite product, we obtain 
$$(q-q^{-1})^2 F_4(z)=\frac{-q\partial_q f(z,q)}{f(z,q)}$$
where $\vartheta_1(z)=q^{\frac{1}{4}}\sin(z)f(z,q).$
As a result we get the identity
$$(q-q^{-1})^2 F_4(z)=
\frac{1}{16}-\frac{q\partial_q \vartheta_1(z)}{4\vartheta_1(z)}
=\frac{1}{16}(\frac{\vartheta_1''(z)}{\vartheta_1(z)}+1).$$

\end{document}